\documentclass[aps,prl,twocolumn,superscriptaddress,showpacs,floatfix]{revtex4-2}

\usepackage{amsmath,amssymb,amsfonts,float,graphics,epsfig,epstopdf,color,verbatim,tabularx,bm,multirow,appendix}
\usepackage[utf8]{inputenc}
\usepackage[T1]{fontenc}
\usepackage{xcolor}
\usepackage{dsfont}
\usepackage{textcomp}
\usepackage{yfonts}
\usepackage{footnote}
\usepackage{bm}
\usepackage{subfigure}
\usepackage{mathrsfs}
\usepackage{graphicx}
\usepackage{verbatim}
\usepackage[colorlinks=true, citecolor=blue, linkcolor=blue, urlcolor=blue]{hyperref}
\usepackage{multirow}
\usepackage{tabularx}
\usepackage{diagbox}
\usepackage{colortbl}
\usepackage{braket}
\usepackage[normalem]{ulem}
\usepackage{tikz}
\usetikzlibrary{calc}
\usetikzlibrary{shapes.multipart}

\newcommand{\comments}[1]{}

\newcommand{\Nc}{N_{\mathrm{c}}}
\newcommand{\scee}
{S^\mathrm{sc}}

\newcommand{\zr}{\mathcal{Z}^{(2)}}
\newcommand{\Lmin}{L_\text{min}}
\newcommand{\ssr}{S^{(2)}}

\newcommand{\lj}[1]{{\color{black}  #1}}

\newcommand{\mhs}[1]{{\color{black}  #1}}

\newcommand{\stkout}[1]{\ifmmode\text{\sout{\ensuremath{#1}}}\else\sout{#1}\fi}
\newcommand{\cx}[1]{{\color{black} #1}}

\makeatletter
\def\l@subsubsection#1#2{}
\makeatother

\begin{document}

% \title{Deconfined quantum criticality lost}
\title{Evolution of entanglement entropy at SU($N$) deconfined quantum critical points}

\author{Menghan Song}
\affiliation{Department of Physics and HK Institute of Quantum Science \& Technology, The University of Hong Kong, Pokfulam Road, Hong Kong}

\author{Jiarui Zhao}
\affiliation{Department of Physics and HK Institute of Quantum Science \& Technology, The University of Hong Kong, Pokfulam Road, Hong Kong}

\author{Meng Cheng}
\affiliation{Department of Physics, Yale University, New Haven, Connecticut 06511-8499, USA}

\author{Cenke Xu}
\affiliation{Department of Physics, University of California, Santa Barbara, CA 93106}

\author{Michael M. Scherer}
\affiliation{Institute for Theoretical Physics III, Ruhr-University Bochum, D-44801 Bochum, Germany}

\author{Lukas Janssen}
\affiliation{Institut f\"ur Theoretische Physik and W\"urzburg-Dresden Cluster of Excellence ct.qmat, TU Dresden, 01062 Dresden, Germany}

\author{Zi Yang Meng}
\email{zymeng@hku.hk}
\affiliation{Department of Physics and HK Institute of Quantum Science \& Technology, The University of Hong Kong, Pokfulam Road, Hong Kong}

\begin{abstract}
Over the past two decades, the enigma of the deconfined quantum critical point (DQCP) has attracted broad attention across the condensed matter, quantum field theory, and high-energy physics communities, as it is expected to offer a new paradigm in theory, experiment, and numerical simulations that goes beyond the Landau-Ginzburg-Wilson framework of symmetry breaking and phase transitions. However, the \cx{nature} of DQCP has been controversial. For instance, in the square-lattice spin-1/2 $J$-$Q$ model, believed to realize the DQCP between N\'eel and valence bond solid states, conflicting results, such as first-order versus continuous transition, and critical exponents incompatible with conformal bootstrap bounds, have been reported. The enigma of DQCP is exemplified in its anomalous logarithmic subleading contribution in its entanglement entropy~(EE), which was discussed in recent studies. In the current work, we demonstrate that similar anomalous logarithmic behavior persists in a class of models analogous to the DQCP.
%\jiarui{To address this issue,} 
We systematically study the quantum EE of square-lattice SU($N$) DQCP spin models.
%, from $N\le 4$ within the $J$-$Q$ model to $N\ge 5$ within the $J_1$-$J_2$ model. 
Based on large-scale quantum Monte Carlo computation of the EE, we show that for a series of $N$ smaller than a critical value,
%$N \le 7$, 
the anomalous logarithmic behavior always exists in the EE, which implies that the previously determined DQCPs in these models do not belong to conformal fixed points. In contrast, 
when $N\ge \Nc$ with a finite $\Nc$ that we evaluate to lie between $7$ and $8$, the DQCPs are consistent with conformal fixed points that can be understood within the Abelian Higgs field theory with $N$ complex components. 
%From the viewpoint of quantum entanglement, our results indicate the realization of a genuine DQCP between N\'eel and valence bond solid phases at finite $N$, and yet 
%explain why 
%\cx{suggest} the SU(2) case is ultimately weakly first order, 
%and pseudocritical, 
%as a consequence of a collision and annihilation of the stable critical fixed point of the $N$-component Abelian Higgs field theory with another, bicritical, fixed point, in agreement with four-loop renormalization group calculations. The experimental relevance of our findings is discussed. 
%\cx{(comment: I think based on just the EE data, we cannot really conclude that the SU(2) case is pseudo critical. We can add a comment in the end which suggests that the pseudo-criticality is a possible explanation.)}

\end{abstract}

\date{\today}
\maketitle

\noindent{\textcolor{blue}{\it Introduction.}---}%
{Over the past two decades, the perplexing enigma of the deconfined quantum critical point (DQCP)~\cite{senthilQuantum2004,sandvikEvidence2007,nahumDeconfined2015,qinDuality2017,wangDeconfined2017,senthilDeconfined2023} has {attracted broad} attention across the communities of condensed matter and quantum materials to quantum field theory and high-energy physics. The DQCP offers a new paradigm in theory beyond the Landau-Ginzburg-Wilson framework of symmetry breaking and phase transitions~\cite{senthilDeconfined2004, senthilQuantum2004, qinDuality2017,wangDeconfined2017,senthilDeconfined2023}, which has inspired fascinating theoretical ideas such as the connection to the 't Hooft anomaly and higher-dimensional symmetry protected topological states~\cite{ashvinsenthil}, emergent symmetry and fractionalized degrees of freedom~\cite{senthilfisher,nahumEmergent2015,maRole2019,sreejithEmergent2019,maDynamics2018}, etc. It has since attracted enormous efforts in numerical simulations~\cite{haradaPossibility2013, sandvikEvidence2007, louvbsneel2009, liuSuperconductivity2019, liaoDiracI2022, shaoQuantum2016, maDynamics2018}, and experiments~\cite{jimenezquantum2021, zayed4spin2017, guoQuantum2020, sunEmergent2021, cuiProximate2023,guoDeconfined2023,nayandqcp}. However, the nature of DQCP have remained highly controversial.} Take the square-lattice SU(2) $J$-$Q$ model~\cite{sandvikEvidence2007} as an example: it was initially believed to realize a continuous quantum phase transition between N\'eel and valence bond solid (VBS) states, but over the years, conflicting results have been reported, such as first-order versus continuous transition~\cite{kuklovDeconfined2008,jiangFrom2008,chenDeconfined2013,demidioDiagnosing2021,maTheory2020,nahumNote2020}, critical exponents that are found to be incompatible with conformal bootstrap bounds
%, assuming SO(5) symmetry without a relevant SO(5) singlet
~\cite{nahumDeconfined2015,shaoQuantum2016,nakayamaNecessary2016,poland}, or possible multi-critical behavior~\cite{zhaoMulticritical2020}. No consensus has been reached to date. 

Similar complications also occur in many more recent DQCP models, such as the fermionic models realizing sequences of transitions from a Dirac semimetal (DSM) through a quantum spin Hall (QSH) insulator to a superconductor (SC)~\cite{liuSuperconductivity2019,liuFermion2023}, or from a DSM through a VBS to an antiferromagnet (AFM)~\cite{liaoDiracI2022,liaoTeaching2023}.
%,
%or from a symmetric DSM through an AFM semimetal to a gapped insulator or superconductor~\cite{liuMetallic2022} \cx{(comment: if Ref.34 observed an intermediate AFM phase with gapless fermion excitations, then it is not really the same situation as the others, I suggest we remove this sentence)} \lj{OK.}.
%
Although the fermionic models have several advantages over the $J$-$Q$~model, e.g., the absence of symmetry-allowed quadruple monopoles and the associated second length scale that corresponds to the breaking of the assumed U(1)~symmetry down to $\mathbb{Z}_4$, incompatible critical exponents persist, and the accumulating numerical results also point towards the absence of a conformal field theory (CFT) of these DQCPs~\cite{liuSuperconductivity2019,wangDoping2021,wangPhases2021,liaoDiracI2022,liuFermion2023,liaoTeaching2023,liuMetallic2022}. 

%Overall, after two decades of efforts, the lattice realizations of the DQCP in its original sense of beyond Landau-Ginzburg-Wilson and yet still critical, with emergent continuous symmetry and fractionalized excitations, are still in ``The Enigma of Arrival,'' and the mystery of their failure of arrival is yet to be explained.
% In fact, the drift in exponents, their conflicts with conformal bootstrap bounds, and the seemingly weakly-first-order behavior appear to be a universal phenomenon of the quantum transition between SU(2) and U(1) or $\mathbb{Z}_4$ breaking phases, pointing towards a common origin. \cx{(comment: suggest remove)}

One clear sign of the perplexity of the DQCP is the anomalous logarithmic subleading contribution to the perimeter law in the finite-size scaling form of R\'enyi entanglement entropy (EE). It is known that for a %real unitary 
CFT in $2+1$ dimensions, the second R\'enyi EE scales as%
~\cite{fradkinEntanglement2006,laflorencieQuantum2016}
\begin{equation}
S^{(2)}_{A}(l_A) = a l_A- s \ln l_A +c+O(1/l_A),
\label{eq:eq2}
\end{equation}
where $l_A$ is the length of the boundary between the entanglement region $A$ and its complement
%the %environment 
$\overline{A}$, $a$ is the coefficient of the perimeter law term, $s$ is the coefficient of the logarithmic correction (log-correction), $c$ is a constant and $O(1/l_A)$ denotes the leading finite-size correction. While the value of $a$ is non-universal, the universal coefficient $s$ of the log-correction depends only on the geometry of the entanglement region $A$ for a given CFT. Crucially, in a CFT when the boundary of $A$ is smooth without any sharp corners, $s$ must vanish. In contrast, if the boundary of $A$ has sharp corners, then in a unitary CFT $s$ is generally positive and its value depends on the opening angles of the corners~\cite{fradkinEntanglement2006,cardyFinite1988,calabreseEntanglement2004}.

%with opening angles $\alpha_i$, then one expects $s=\sum_i s(\alpha_i)$. The function $s(\alpha)$ is characteristic for the respective CFT~\cite{fradkinEntanglement2006,cardyFinite1988,calabreseEntanglement2004}.
%realizes a regular polygon, the value of $s$ is proportional to the number of corners of the polygon, with a positive constant of proportionality that is characteristic for the respective CFT~\cite{Casini2012}.
%
It is the goal of this work to show that the anomalous logarithmic subleading contribution, in particular a non-zero value of $s$ for entanglement regions with smooth boundary, is actually very ubiquitous, in a series of models that can be viewed as SU($N$) generalizations of the DQCP. Vanishing of the anomalous log-correction determines the critical value $N_c$ of $N$, above which the EE measurement is consistent with the expectations of CFTs.

\begin{figure}[tb!]
\includegraphics[width=\columnwidth]{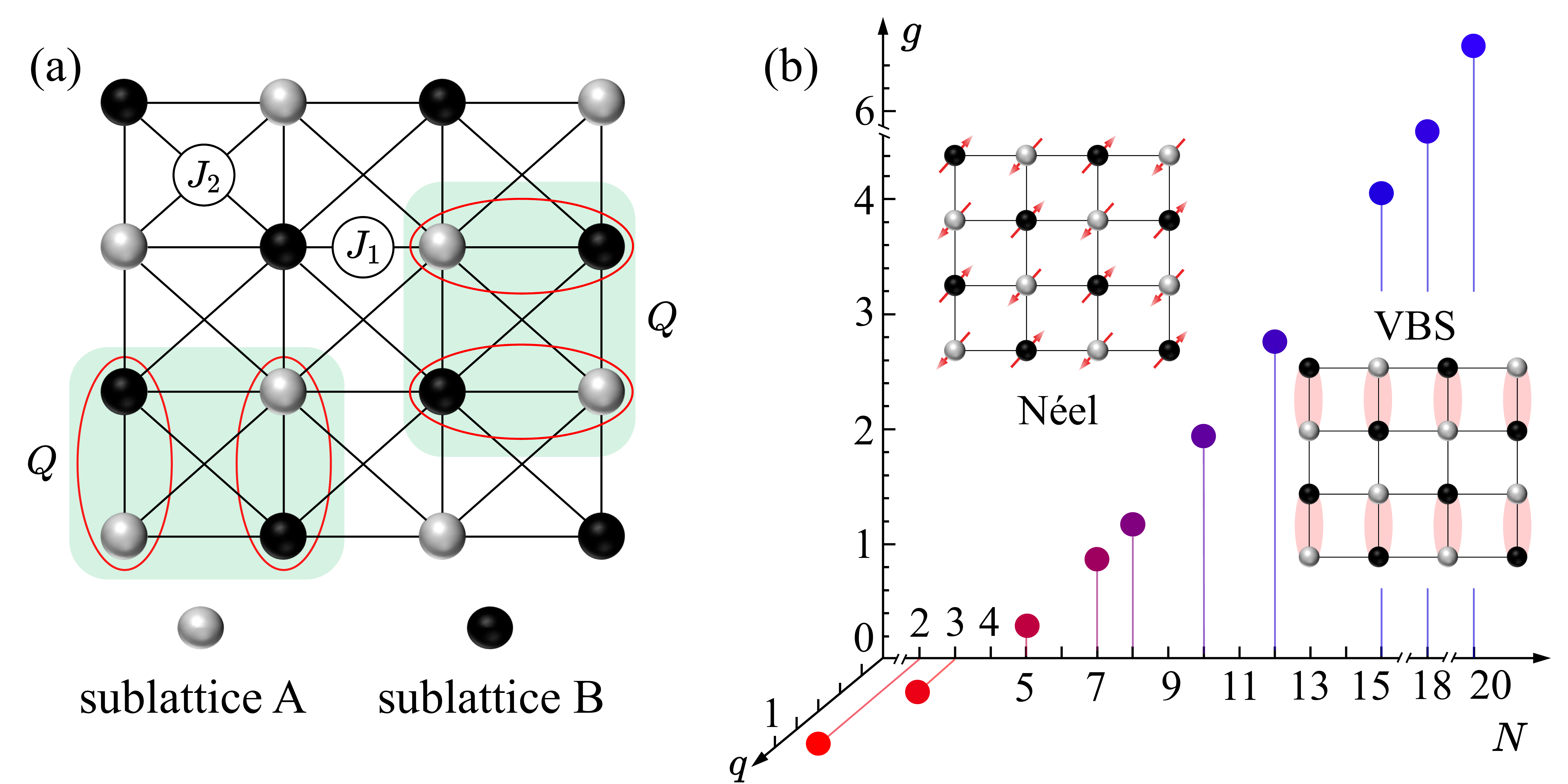}
\caption{\textbf{SU($\boldsymbol N$) $\boldsymbol{J_1}$-$\boldsymbol{J_2}$-$\boldsymbol{Q}$ model and its phase diagram.} 
(a)~$J_1$-$J_2$-$Q$ model on square lattice with white (black) sites representing sublattice A (B). Solid lines correspond to nearest-neighbor antiferromagnetic exchange $J_1$ and next-nearest-neighbor ferromagnetic exchange $J_2$. Green shaded squares indicate four-spin ring exchange $Q$.
(b)~Phase diagram as function of $g=J_2/J_1$, $q=Q/(Q+J_1)$, and $N$. Colored dots indicate transition points, at which we analyze the finite-size scaling behavior of the EE. When $N\le4$, the transition is tuned by $q$ between N\'eel at small $Q$ and VBS at large $Q$. When $N>4$, the transition is tuned by $g$ between VBS at small $J_2$ and N\'eel at large $J_2$. 
%
% We find the transitions for \cx{$N \leq 5$} are not compatible with unitary CFTs, showing weakly-first-order transitions denoted by the red-colored crosses. 
% In contrast, the models with large enough $N$ are compatible with unitary CFTs and continuous behavior, indicated by the blue-colored dots. 
}
\label{fig:fig1}
\end{figure}

\noindent{\textcolor{blue}{\it Main result.}---}%
Our new approach of analyzing DQCP is to systematically investigate the scaling of the second-order R\'enyi EE upon partitioning into subregions with smooth boundaries as well as subregions with corners~\cite{casiniUniversal2006}.
As a proof of concept, we choose the square-lattice SU($N$) DQCP spin model~\cite{kaulQuantum2012,kaulBridging2013,kaulLattice2012,blockFate2013} from $N=2,3$ (the $J$-$Q$ model) to \mhs{$N=5,7,8,10,12,15,18,20$ (the $J_1$-$J_2$ model)}, see Fig.~\ref{fig:fig1}(a). Using the non-equilibrium incremental quantum Monte Carlo (QMC) algorithm to measure the EE~\cite{albaOut2017, jonNoneq2020, zhaoScaling2022,zhaoMeasuring2022}, we show that for $N=2,3,5,7$, the previously determined DQCPs all show 
a finite log-correction, for subregions with {\it smooth} boundaries.
%as well as entanglement boundaries including corners. 
These DQCPs are therefore incompatible with  CFT descriptions, and are most likely weakly first-order.
%\jiarui{a positive logarithmic corrections to the leading perimeter law term of EE scaling with both smooth and corner entanglement boundaries, which is incompatible with \lj{unitary} CFTs}.
%
%As a consequence, these DQCPs cannot correspond to \lj{direct} continuous transitions \lj{described by unitary CFTs, and are most likely weakly first order.}
%and are most likely weakly first order \meng{, as the possible explanation of the positive log corrections comes from the Goldstone modes of the remaining order, either N\'eel or the breaking of emergent symmetry~\cite{song2024extracting,dengDiagnosing2024}, at the transition point.}\jiarui{[Jiarui: I think our results only suggest these DQCPs are not CFTs and can't preclude the possibility of continuous quantum phase transitions?]} \meng{How about now?}.
%
%
In contrast, when $N \geq 8$, the EE scaling with smooth entanglement boundaries for the DQCPs no longer has an obvious logarithmic subleading correction, and they are compatible with continuous phase transitions.
%are fully compatible with constraints from unitary CFTs 
%and are most likely continuous. 
\mhs{This is further supported by the EE scaling at $N=8,10,...,18,20$ for regions with corners, which shows a logarithmic correction with $s>0$.}
\mhs{In fact,  we find that our numerically extracted value of $s$ for $N\ge 8$} is reasonably consistent with the expectations from the Abelian Higgs theory in the large-$N$ limit~\cite{senthilQuantum2004, sachdevQuantum2008, kaulBridging2013, nahumDeconfined2015},
which features unitary conformal fixed points~\cite{irkhinExpansion1996, kaulQuantum2008, ihrigAbelian2019}. Thus our results suggest the existence of a finite critical $\Nc$ above which the DQCP becomes continuous.
{Based on the behavior of EE with smooth boundary, our numerical results suggest that $\Nc$ lies between $7$ and $8$.} 

Distinguishing a weakly-first-order transition from a truly continuous one is a challenging numerical task when using conventional local observables and their correlation functions.
In our work, this difficulty is overcomed by studying the EE, which is a non-local observable and can reveal subtle structures in quantum many-body wavefunctions beyond conventional measurements%
%based on local observables and their correlation functions
~\cite{calabreseEntanglement2004,fradkinEntanglement2006,levinDetecting2006,kitaevTopological2006,laflorencieQuantum2016,zhaoMeasuring2022}. The log-coefficient of the EE has to satisfy the $\mathit{positivity} $ requirement for a unitary CFT~\cite{Casini2012}.
Our results support the realization of a true DQCP between N\'eel and VBS phases at finite but large $N$, and allow
%
%It allows 
us to demonstrate the absence of a conformal fixed point for $N=2,3,5,7$.
%
%Our results prove the realization of a true DQCP between N\'eel and VBS phases at finite but large $N$, and yet explain why the SU(2) case is \emph{not} a true DQCP, but a result of FP annihilation, leading to pseudocritical behavior and a complex nearby CFT~\cite{maTheory2020,nahumNote2020,ihrigAbelian2019}. \mc{How do the EE results prove these statements? At best, we can only say that the EE results support this scenario...}
%\meng{Meng, can you help with revising here? Or, we can just go back to the weakly first order transition as our statement.}
%
%This explains all the previous difficulties of the violation of the bootstrap bound~\cite{nakayamaNecessary2016}, the drifting of the critical exponents~\cite{nahumDeconfined2015}, and the indications for weakly-first-order transitions~\cite{wangPhases2021,demidioDiagnosing2021}, for $N \leq 7$. 

%\mc{I think we are overstating in the last sentence, which I commented out. }
%

%Such understanding also naturally explains why, experimentally, both in the VBS-AFM DQCP transition in the quantum magnet SrCu$_2$(BO$_3$)$_2$~\cite{zayed4spin2017,guoQuantum2020, jimenezquantum2021,cuiProximate2023}, and the QSH-SC superconducting quantum criticality in monolayer WTe$_2$~\cite{songUnconventional2023}, the systems either exhibit first-order transitions or intermediate phases, defying the original proposal for an SU(2) DQCP. 

\noindent{\textcolor{blue}{\it Model and phase diagram.}---}%
We study the SU($N$) spin model defined in a Hilbert space of $N$ local states (colors) at each site of the square lattice~\cite{kaulQuantum2012,kaulBridging2013,kaulLattice2012,blockFate2013}, as shown in Fig.~\ref{fig:fig1}(a). We assume SU($N$) spins in the fundamental representation on sublattice $A$ and in the conjugate representation on sublattice $B$, i.e., $|\alpha\rangle_A \to U_{\alpha,\beta}|\beta\rangle_A$, $|\alpha\rangle_B\to U^{*}_{\alpha,\beta}|\beta\rangle_B$, with the state $\sum_{\alpha}|\alpha\rangle_A |\alpha\rangle_B$ an SU($N$) singlet~\cite{Ianlarge-n1985,read1989}. The Hamiltonian reads
\begin{equation}
	H=-\frac{J_{1}}{N} \sum_{\langle i j\rangle} P_{i j}-\frac{J_{2}}{N} \sum_{\langle\langle i j\rangle\rangle} \Pi_{i j}-\frac{Q}{N^2} \sum_{\langle i,j\rangle, \langle k,l \rangle} P_{i j}P_{k l},
	\label{eq:eq1}
\end{equation}
where the $J_1$ term is the SU($N$) generalization of the nearest-neighbor antiferromagnetic interaction, as $P_{ij}$ is defined as the projection operator onto the SU($N$) singlet between a pair of spins $i$ and $j$ on different sublattices, and the $J_2$ term is the SU($N$) generalization of the next-nearest-neighbor ferromagnetic interaction, as $\Pi_{ij}$ is the permutation operator acting between sites having the same representation on the same sublattice, i.e., $\Pi_{ij}|\alpha\beta\rangle=|\beta\alpha\rangle$. We also add a four-spin ring exchange term $Q$ for the $N=2,3,4$ cases, where $\langle i,j\rangle, \langle k,l \rangle$ are spin pairs located on adjacent corners of a 4-site plaquette, see Fig.~\ref{fig:fig1}(a). This term preserves the translational and rotational symmetries of the square lattice, and was found to stabilize a VBS state with $\mathbb{Z}_4$ symmetry-breaking at large $Q$~\cite{louvbsneel2009,kaulLattice2012,haradaPossibility2013}. We compute the second-order R\'enyi EE of the model in Eq.~\eqref{eq:eq1} with QMC on lattices with linear sizes $L=8,12,16,...,40$. \mhs{We keep the inverse temperature $\beta \equiv 1/T =L$ at $N=2$, $\beta=8L$ for $N=18,20$ and $\beta=4L$ for other intermediate $N$ values to circumvent thermal pollution (see SM~\cite{suppl} for detailed analysis).}

The phase diagram of Eq.~\eqref{eq:eq1}, spanned by the axes of $q=\frac{Q}{J_1+Q}$, $g=\frac{J_2}{J_1}$, and $N$, is shown in Fig.~\ref{fig:fig1}(b). It is consistent with previous QMC works~\cite{louvbsneel2009,haradaPossibility2013,BeachcontinuousN2009,kaulQuantum2012,kaulBridging2013,kaulLattice2012,blockFate2013}. At $N=2,3,4$, a transition between N\'eel and VBS state can be induced upon tuning $q=\frac{Q}{J_1+Q}$ for fixed $J_2=0$~\cite{sandvikEvidence2007,Melko2008,jiangFrom2008,louvbsneel2009,Kaul2011}. 
For $N \geq 5$, the $J_1$-only model for $J_2 = Q = 0$ already has a VBS ground state~\cite{readLarge1991,assaadPhase2005,BeachcontinuousN2009}, and a N\'eel-VBS transition can be induced by tuning $g = J_2/J_1$ for fixed $Q=0$~\cite{kaulLattice2012,haradaPossibility2013,BeachcontinuousN2009}. 
In the SM~\cite{suppl}, we show that our critical couplings $q_c$ and $g_c$ agree with those in the literature~\cite{sandvikEvidence2007, louvbsneel2009, jiangFrom2008, Melko2008, Kaul2011, kaulQuantum2012, kaulBridging2013, kaulLattice2012, blockFate2013}. We also determine the corresponding critical exponents at a few representative values of~$N$.
%
%According to the previous works~\cite{kaulQuantum2012,kaulBridging2013,kaulLattice2012,blockFate2013}, the transition points along the axis of $g=J_2/J_1$ are also known upto $N=12$, and we have further determined the transition points for $N=15,20,30$ and the corresponding critical exponents therein as shown in the SM~\cite{suppl}.

\noindent{\textcolor{blue}{\it Finite-size scaling of EE.}---}% 
As reviewed earlier, the subleading corrections to the EE in a CFT needs to satisfy nontrivial conditions~\cite{casiniUniversal2006,casiniEntanglement2007,cardyFinite1988,calabreseEntanglement2004,fradkinEntanglement2006}. 
We now turn to the EE measurements of the transitions in the phase diagram of Fig.~\ref{fig:fig1}(b). To this end, a non-equilibrium incremental QMC algorithm is developed ~\cite{albaOut2017,jonNoneq2020,zhaoScaling2022,zhaoMeasuring2022} for the SU($N$) spin model. Details of the implementation are given in the SM~\cite{suppl}. Here we only mention that to compute the second-order R\'enyi EE $S^{(2)}_A$ for quantum spin systems, there are many previous attempts based on the swap operator and its extensions~\cite{kallin2011anomalies,hastings2010measuring,humeniuk2012quantum,helmes2014entanglement,kulchytskyy2015detecting,isakovTopological2011,laflorencieQuantum2016} and the data quality is always a serious issue when approaching large system sizes for extracting the subleading universal scaling coefficients. This problem has been greatly relieved by the incremental algorithm, which converts the R\'enyi EE into the free energy difference between partition functions on two different manifolds, with the help of Jarzynski equality~\cite{Jarzynski1997, albaOut2017, jonNoneq2020} and the incremental trick~\cite{zhaoScaling2022}. Controlled EE results, including the log-coefficient of the EE inside the N\'eel phase of the antiferromagnetic Heisenberg model and at its (2+1)D O(3) quantum critical points~\cite{jonNoneq2020,zhaoMeasuring2022,zhaoScaling2022,songQuantum2023,song2024extracting} and the topological EE inside the Kagome $\mathbb{Z}_2$ quantum spin liquid~\cite{zhaoMeasuring2022}, have been obtained. 
We note the latest developments, realizing the EE as an exponential observable~\cite{zhangIntegral2023} and simpler incremental approaches without non-equilibrium process~\cite{liaoControllable2023,liaoTeaching2023,zhouIncremental2024}, have been put forward~\cite{liaoControllable2023,liaoTeaching2023,zhangIntegral2023,zhouIncremental2024}. 

The scaling of the EE for a quantum critical point of a 2D lattice model, described by a CFT, is given in Eq.~\eqref{eq:eq2}.
%
%\begin{equation}
%S^{(2)}_{A}(l_A) = a l_A- s \ln l_A +c+O(1/l_A),
%\label{eq:eq2}
%\end{equation}
%
%\mms{This is already Eq. (1).}
%where $l_A$ is the length of the boundary between the entanglement region $A$, and the environment $\overline{A}$, $a$ is the coefficient of the perimeter law term, $s$ is the coefficient of the logarithm correction (log-correction), $c$ is a constant and $O(1/l_A)$ denotes the leading finite-size correction.
%
%It is essential to differentiate entanglement regions with smooth boundaries and those with sharp corners. If the subregion $A$ does not involve any corners [e.g., the bipartition shown in the inset of Fig.~\ref{fig:fig2}(a)], one expects a vanishing log-correction $s=0$ if the critical point is described by a CFT~\cite{helmes2014entanglement,song2024extracting} and the scaling form in Eq.~\eqref{eq:eq2} reduces to $S^{(2)}_A(l_A) = al_A+c+O(1/l_A)$. By contrast, if the subregion $A$ involves sharp corners with angle $\alpha_i$ [e.g., the cut shown in the inset of Fig.~\ref{fig:fig3}(a), which involves four sharp corners with angle $\alpha=\pi/2$], one expects a finite log-correction of the form $s=\sum_i s(\alpha_i)$. 
%
%
%realized on the $\pi$-flux square lattice with two spin flavors and two Dirac cones in the Brillouin zone, with four $\pi/2$ corners of region $A$~\cite{helmesUniversal2016,liaoTeaching2023}, 
%
%
In a CFT, the coefficient $s$ can be written as $s=\sum_i s(\alpha_i)$,
where $\alpha_i$ is the opening angle of the $i$-th corner on the boundary of the region $A$. Here $s(\alpha)$ is a universal quantity for a CFT~\cite{fradkinEntanglement2006,cardyFinite1988,calabreseEntanglement2004}, satisfying a number of nontrivial constraints. For our purpose, the following two conditions are the most relevant~\cite{Casini2012,buenoBounds2016,buenoUniversality2015}:
\begin{enumerate}
    \item In a CFT we must have $s(\pi)=0$ . This is equivalent to the statement that smooth entanglement cuts should have no log-correction.
    \item In a unitary CFT $s(\alpha)\geq s(\pi)=0$ for $\alpha\in(0,\pi]$.
\end{enumerate}
The corner contribution has previously been numerically and/or analytically computed for different CFTs. For example, it is known that
$s(\pi/2) = 0.01496$ for a single (2+1)D Dirac fermion CFT~\cite{helmesUniversal2016,liaoTeaching2023},
$s(\pi/2) = 0.0064$ for a single real free boson~\cite{casiniUniversal2006},
and $4s(\pi/2)=0.081(4)$ for a square region at the $(2+1)$D O(3) transition~\cite{zhaoMeasuring2022,zhaoScaling2022,songQuantum2023,kallinCorner2014,song2024extracting}.
%between N\'eel state and paramagnetic state in spin-1/2 square-lattice model.
%
%Moreover, the sign of $s$ in Eq.~\eqref{eq:eq2} satisfies an important $\mathit{positivity}$ condition: In a \emph{unitary} CFT, one must have $s(\alpha)\geq s(\pi)=0$ for all $\alpha\in [0,\pi]$~\cite{Casini2012,buenoBounds2016,buenoUniversality2015}.
%
In addition, a spontaneous-symmetry-breaking (SSB) phase with Goldstone modes is expected to exhibit a scaling form of the EE analogous to Eq.~\eqref{eq:eq2}, with an additional contribution $s_\text{G} = - n_\text{G}/2$ to the coefficient $s$ of the log-correction, where $n_\text{G}$ corresponds to the number of Goldstone modes~\cite{metlitski2015entanglement}.

\begin{figure*}[tp!]
\includegraphics[width=\textwidth]{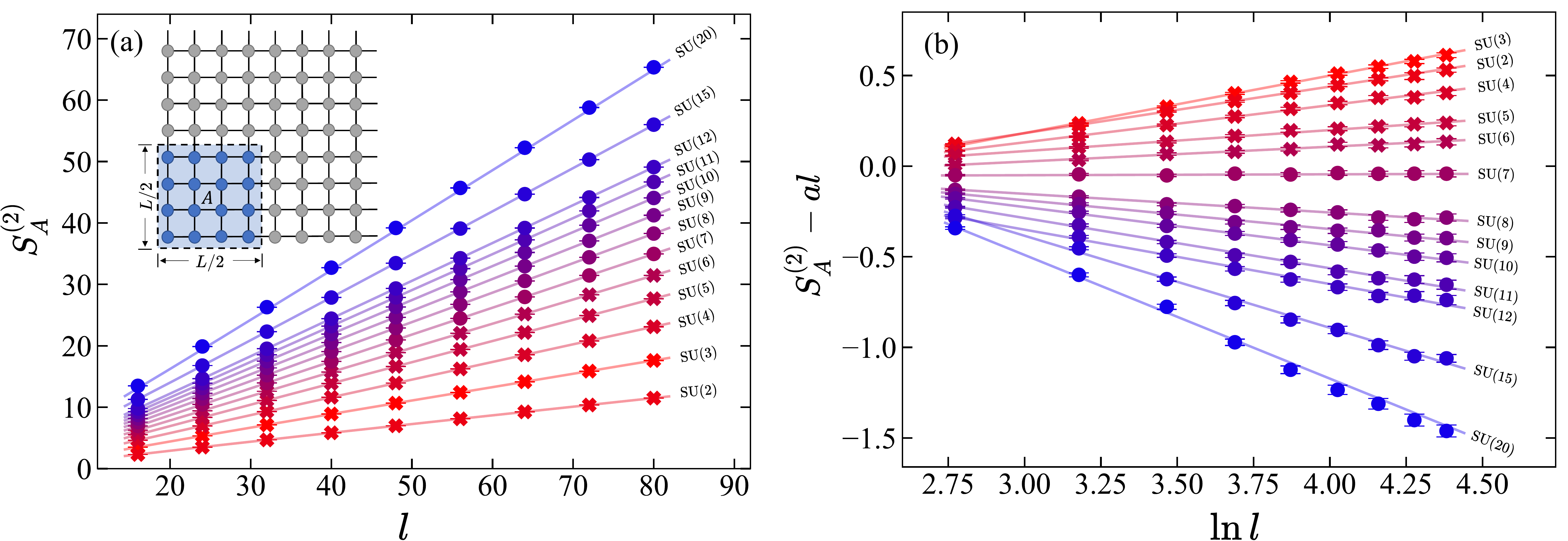}
\caption{%
\textbf{Second-order R\'enyi EE and its scaling behavior at the SU($\boldsymbol{N}$) DQCPs with smooth boundaries.}
(a)~EE $S^{(2)}_{A}$ as function of boundary length $l_A=2L$ of the entanglement region $A$ with smooth boundaries for different values of $N$, with the partitioning shown in the inset. The \lj{perimeter} law behavior $\propto a l_A$ becomes more prominent for increasing $N$.
(b)~EE with \lj{perimeter} law contribution subtracted, i.e., $S^{(2)}_{A}-al_A-\mathrm{constant}$, as a function of $\ln l_A$ for different values of $N$. The slopes of these curves reflect the coefficient $s$ of the logarithmic term in Eq.~\eqref{eq:eq2}. 
%with a positive slope corresponding to a negative value of $s$, which is forbidden for a quantum critical point described by a CFT. 
\cx{For $N \leq 7$, there is a finite nonzero $\ln L$ subleading correction to the perimeter law;} while as $N$ increases over $8$, the slope decreases and eventually vanishes at sufficiently large $N$. \cx{In fact, for $N > 8$, the subleading correction fits better with the form $1/L$ rather than $\ln L$, as we discuss quantitatively with the $\chi^2/k$ value and the subtracted EE in the supplementary material.} For illustration purposes, the fitting in (b) starts from $L_\text{min}=16$ at all $N$ values.
(c)~Finite-size drift of fitted $s$ as a function of $1/L_\text{min}$, where $L_\text{min}$ corresponds to the smallest retained system size in the fitting process. 
%For small $N$, $s$ remains finite and negative upon decreasing $1/L_\text{min}$. However, at $N\geq 8$, $s$ gradually drifts to zero upon decreasing $1/L_\text{min}$, permitting a CFT description.
%
%Colors of data points for different $N$ indicate the corresponding value of the log-coefficient $s$ with red (blue) representing $s<0$ ($s = 0$).
}
\label{fig:fig2}
\end{figure*}

\noindent{\textcolor{blue}{\it EE with smooth boundaries.}---}%
Our QMC-obtained EE for the Hamiltonian in Eq.~\eqref{eq:eq1} with smooth bipartition (or equivalently with $\alpha=\pi$ corners) are shown in Fig.~\ref{fig:fig2}.
%For the EE computation, we have simulated the square lattice with linear system sizes of $L=8,12,\cdots 40$ with replicas and the inverse temperature $\beta=L$. 
Since the entanglement region $A$ is of the size $L \times L/2$, the boundary length of $A$ is $l_A=2L$. We plot $S^{(2)}_A(l_A)$ as a function of $l_A$ for each $N$ at its corresponding putative DQCP and fit a functional form according to Eq.~\eqref{eq:eq2}.

As shown in Fig.~\ref{fig:fig2}(a), the obtained $S^{(2)}_A$ for all $N$ values are dominated by the \lj{perimeter} law scaling, i.e., when $l_A$ becomes large, a linear term in $S^{(2)}_A$ manifests. However, a clear difference appears once we subtract the \lj{perimeter} law contribution from the data. In Fig.~\ref{fig:fig2}(b), we plot $S^{(2)}_A - a l_A$ versus $\ln l_A$. The slope of these curves reveals the values of $s$ %and, even more importantly, the signs of $s$ in Eq.~\eqref{eq:eq2} 
for different $N$. One sees that for the cases of SU(2), SU(3), SU(5), and SU(7), we have a finite log-coefficient $s<0$ [revealed by a positive slope in Fig.~\ref{fig:fig2}(b)], violating the equality $s(\pi)=0$. Therefore, our observation of a finite $s(\pi)$ here shows that the putative DQCPs for small $N$, e.g., $N = 2,3,5,7$, are incompatible with a CFT. Figure~\ref{fig:fig2}(c) demonstrates the finite-size analysis for the fitted~$s$. We can see the $s$ values for SU(3), SU(5) and SU(7) are robust as one increasing the smallest retained system size $L_\text{min}$ in the fitting process. Only from SU(8) on, the value of $s$ becomes close to zero $s\approx 0$ as one increases $L_\text{min}$. 

% Exactly the same behavior has been observed in the SU(2) $J$-$Q_3$ model~\cite{zhaoScaling2022} and the other spin-1/2 fermion DQCP models~\cite{liuFermion2023,liaoTeaching2023}. It is interesting to notice that $s=0.25(1)$ for our SU(2) $J$-$Q$ model matches quantitatively with the previously determined value $s=0.25(1)$ for the SU(2) $J$-$Q_3$ model with six-spin $Q_3$ term~\cite{zhaoScaling2022}. These results, combining the observation of the negative log-correction in the EE and disorder operators in the same spin and fermion DQCP models~\cite{wangScaling2021,zhaoScaling2022,liuFermion2023,liaoTeaching2023}, deliver a clear message that all these SU(2) DQCP models are unanimously incompatible with a unitary CFT description.

{The $N$ dependence of the EE scaling exhibited in Figs.~\ref{fig:fig2}(b) and \ref{fig:fig2}(c) has important implications for the fate of SU$(N)$ DQCPs. As one increases $N$ in the SU($N$) model, there is a clear change in the nature of the transition, as indicated by the fitted $s$ shown in Fig~\ref{fig:fig2}(c). \cx{A comprehensive fitting quality analysis is presented in the SM~\cite{suppl}, by comparing the fitting with $\ln L$ and $1/L$ finite size correction, using the $\chi^2/k$ value, as well as the ``subtracted EE'' devised in Ref.~\cite{song2024extracting}.}
%
%
%which shows that the subleading corrections to the EE scaling for $N \leq 7$ are best fitted assuming a finite log-correction, while for large $N$ they are best fitted 
%assuming only residual finite-size corrections 
%\cx{with regular $1/L$ finite size effect and vanishing log-correction}.}
%
For $N=2,3,5,7$, $s$ is found to be finite. 
By contrast, for $N=10, 12, 15, 18, 20$, $s$ vanishes in the thermodynamic limit. Therefore, the behavior of EE at these transitions is compatible with CFTs.
SU(8) represents a boundary case in which the fitted $s$ of smooth cut becomes indistinguishable from zero within the error bar, and the subleading correction to the perimeter law fits equally well with $\ln L$ and $1/L$ (please refer to the SM~\cite{suppl}). 
%(comment: I rephrased this sentence please check)}
%
The SU($N \geq 8$) N\'eel-to-VBS transitions are, therefore, candidates for genuine DQCPs in the original sense, i.e., continuous quantum phase transitions between two different SSB phases, described by CFTs.
This suggests the existence of a finite critical value $\Nc$, above which the transition becomes continuous.
Our numerical data shown in Fig.~\ref{fig:fig2}(c) for smooth boundary, together with the extended analysis shown in Figs.~\ref{fig:sm8} and \ref{fig:sm9} of the SM~\cite{suppl}, suggest that $\Nc$ lies between $7$ and $8$.
%
%Our numerical data shown in Fig.~\ref{fig:fig2}(c) indicate that $\Nc$ lies between $8$ and $10$.

%$N \leq 7$, gradually changes from a finite and negative value at  to a small one that is difficult to distinguish numerically from zero at large system sizes for SU(8). Here, one sees the DQCPs from $N\gtrsim 8$ become compatible with a unitary CFT as $s(\pi)=0$ and only from this value of $N$ on. For the SU($N \geq 10$) N\'eel-to-VBS transitions, $s(\pi)$ quickly vanishes in the thermodynamic limit, we obtain candidates for genuine DQCP in the original sense, i.e., a continuous phase transition between two different spontaneously symmetry-breaking phases, described by a unitary CFT.

%In fact, for a first-order transition, vanishing of $s$ for several consecutive values of $N$ requires fine tuning, and is therefore unlikely. \mc{MC: this sentence is a bit strange.} 
%
%A natural explanation for the vanishing of $s$ for $N \gtrsim 8$ is that the transition becomes truely continuous when $N$ is above a certain critical value of $N_c$.
%Our numerical data shown in Fig.~\ref{fig:fig2}(c) suggest that $\Nc$ lies between $8$ and $10$. 
%%which agrees with the field-theoretical estimate within the error bars.
%}

%\mc{MC: I moved the AH field theory discussion to the end of the paper.}
%\ljcomment{Good!}

\begin{figure*}[tb!]
\includegraphics[width=\textwidth]{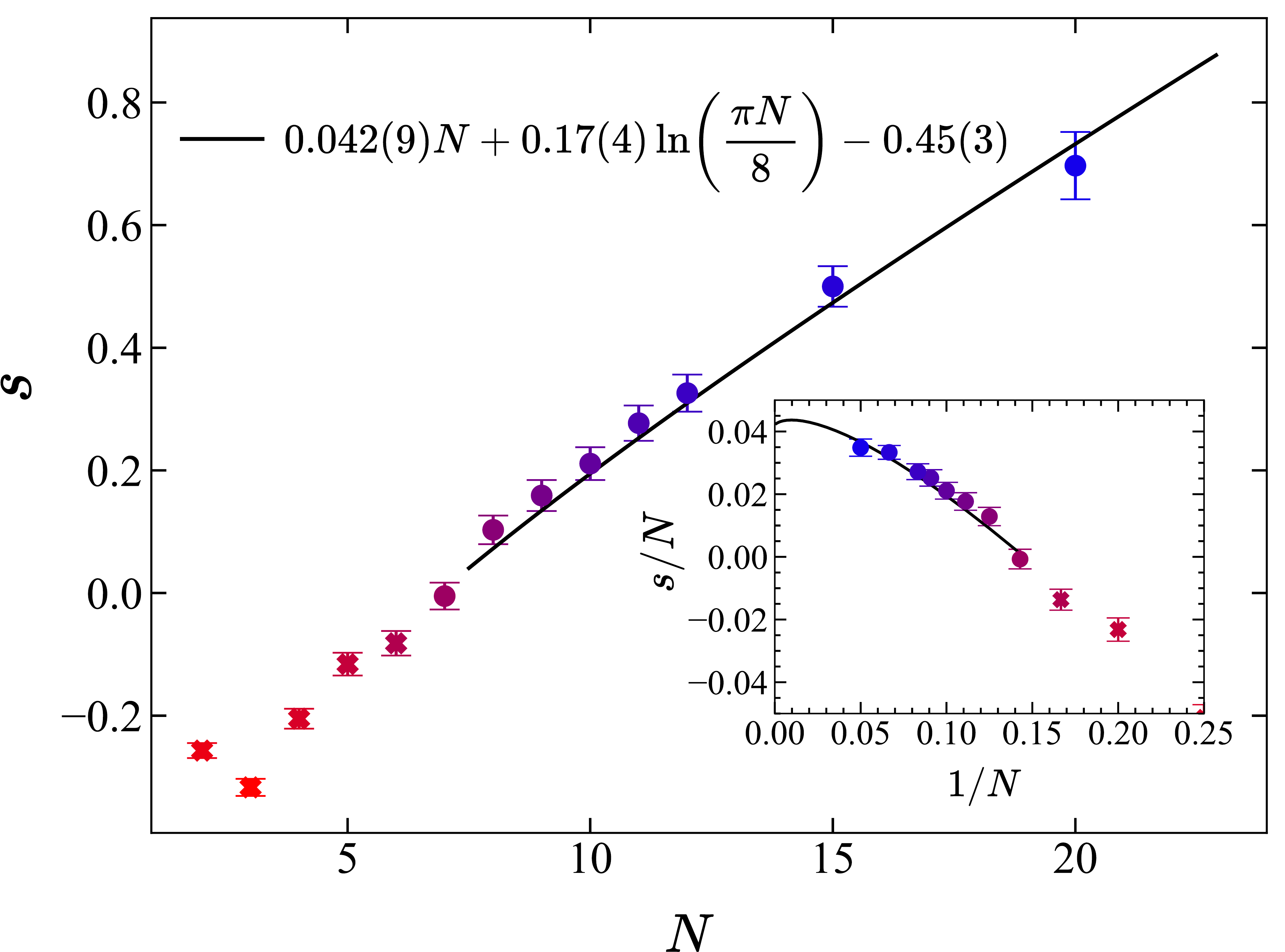}
\caption{\textbf{Second-order R\'enyi EE and its scaling behavior at the SU($\boldsymbol{N}$) DQCPs with corner cuts.} \mhs{(a) EE $S^{(2)}_{A}$ as a function of boundary length $l_A=2L$ of the entanglement region $A$ with corner cuts at SU(3). The upper inset demonstrates the entanglement region $A$ with four $\pi/2$ corners. The lower inset reflects its negative log-correction $s<0$, which we attribute to the contribution from the smooth part of the boundary. (b) Subtracted corner entanglement entropy $\scee$ as a function of $\ln l_A$ at $N>N_c$. $\scee$ is defined as the difference between $S^{(2)}_A$ of the smooth (red) and corner (blue) regions that have the same boundary length $l_A=2L$, as shown in the inset. The slope of the linear fitting equals the log-coefficient $s$ which purely comes from four $\pi/2$ corners and monotonically increases against $N$. For $N>N_c$, $s>0$ for all cases studied in this work, consistent with the CFT constraint. The linear fitting in (b) starts from $\Lmin=16$ for all $N$ values. Panel (c) demonstrates the change of fitted $s$ against the smallest system sizes retained in the fitting process. Benefiting from the linear scaling of $\scee$ against $\ln l_A$, the fitted $s$ does not drift much as increasing $\Lmin$. The dashed line denotes the averaged $s$ values among all $\Lmin$ cases.}}
\label{fig:fig3}
\end{figure*}

\noindent{{\textcolor{blue}{\it EE with sharp corners.}}---}%
We also analyze the subleading contribution to EE for subregions with corners, especially for $N > \Nc$ {where continuous transitions are expected and the corner coefficient is expected to be universal}. 

\mhs{Figure~\ref{fig:fig3}(a) presents the scaling of R\'enyi EE for a subregion $A$ with four $\alpha=\pi/2$ corners, as depicted in the upper inset, at the SU(3) DQCP as a representative case for $N<N_c$. 
As shown in the lower inset of Fig.~\ref{fig:fig3}(a), the EE with four $\alpha=\pi/2$ corners of the SU(3) DQCP clearly shows a negative log-coefficient, $s=-0.35(2)$. Interestingly, the logarithmic coefficient with four sharp $\pi/2$ corners is close to those we have obtained with smooth boundaries, $s(\pi) = - 0.34(2)$, as depicted in Fig.~\ref{fig:fig2}(b,c). This suggests that the observed log-corrections with corners for small $N$ are in fact inherited from the smooth boundary case. Similar behavior has also been observed in the SU(2) $J$-$Q_2$ and $J$-$Q_3$ model~\cite{song2024extracting,zhaoScaling2022,dengDiagnosing2024} and fermion DQCP models~\cite{wangScaling2021,zhaoScaling2022,liuFermion2023,liaoTeaching2023}. The corner contribution in these cases is too small to be numerically detectable compared to the large negative $s(\pi)$ of smooth boundaries.}

\begin{figure}[tb!]
\includegraphics[width=0.85\columnwidth]{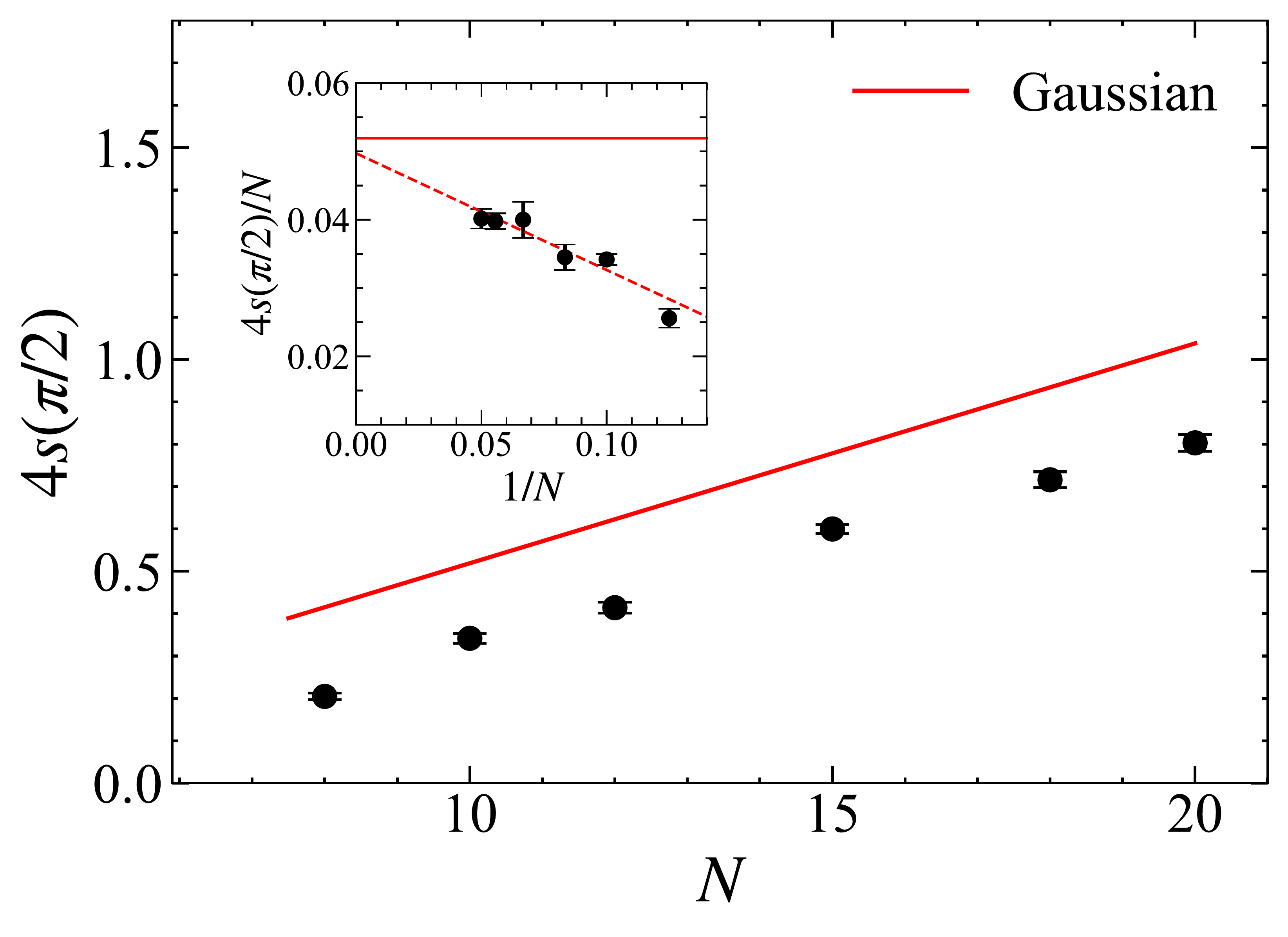}
\caption{\textbf{Fitted log-coefficient from four $\pi/2$ corners at large-$N$.} \mhs{The main panel \lj{shows $4s(\pi/2)$ as a function of $N$ with the red line indicating the corresponding Gaussian value.} %that in a Gaussian theory with $N$ complex components, i.e., 
%$0.0064\times 2\times N\times 4$. 
The black dots are the averaged $s$ values among all $\Lmin$ in Fig.~\ref{fig:fig3}(c), with the error bar denoting the standard deviation.} \lj{The inset shows $4s(\pi/2)/N$ as a function of $1/N$, together with a linear fit (dashed line), yielding $4s(\pi/2)/N = 0.050(2) - 0.17(2)/N + O(1/N^2)$, which agrees with the Gaussian value (solid line) for $N \to \infty$.}
%The red dashed line in the inset is a linear fitting of $4s(\pi/2)/N$ \lj{as function of $1/N$} which goes to the Gaussian value for a complex scalar (horizontal red line) at $N\to \infty$ The slope of the dashed line represents the subleading constant correction to the free Gaussian theory which is fitted to be -0.17(2).}
}
\label{fig:fig4}
\end{figure}

\mhs{At $N\ge \Nc$, we observe from Fig.~\ref{fig:fig2} that $s(\pi)$ vanishes as expected from CFT predictions. In this case, the logarithmic correction is caused by four sharp $\pi/2$ corners. To extract the subleading terms in EE precisely, we utilize a recently developed algorithm~\cite{liao2024extracting} to measure the subtracted corner entanglement entropy ($\scee$), defined as the difference between the EEs of subregions with the same boundary length for smooth and cornered boundaries, i.e., red and blue regions in the inset of Fig.~\ref{fig:fig3}(b) respectively (see Sec.~\ref{algorithms} in SM) {in one Monte Carlo simulation. With this method,} the leading perimeter law contribution is automatically canceled out, and $\scee$ scales as $\scee = s\ln l_A + c +O(1/l_A)$ according to Eq.~\eqref{eq:eq2}. Here, $s$ comes entirely from the corner contributions, and a linear fit of $\scee$ against $\ln l_A$ is stable enough to extract subleading logarithmic coefficient.}

\mhs{Figure~\ref{fig:fig3}(b) presents the $\scee$ data for $N\ge \Nc$ up to $N=20$. The slope of the linear fitting equals the log-coefficient from four $\pi/2$ corners, that is, $4\times s(\pi/2)$, and its drift against $\Lmin$ is shown in Fig.~\ref{fig:fig3}(c). For all $N\ge \Nc$ investigated, the $s(\pi/2)$ values are compatible with the \emph{positivity} constraint. Importantly, $s(\pi/2)$ values are also consistent with the theoretical expectation from $N$-component Abelian-Higgs and non-compact CP$^{N-1}$ field theories at leading order, which have been suggested as continuum descriptions of the SU($N$) DQCPs~\cite{dyerScaling2015, kaulLattice2012, senthilDeconfined2023}:
In the large-$N$ limit, these theories are weakly coupled~\cite{ihrigAbelian2019}.
%As the non-compact CP$^{N-1}$ field theories in large$-N$ are weakly interacting, 
\lj{We therefore expect} the leading contribution to the log-coefficient \lj{at large $N$ to }%should therefore
be given by the corresponding Gaussian theory of the scalar bosons~\cite{Klebanov:2011td}. For the Abelian-Higgs model with $N$ complex components, this implies $s$ is linear against $N$ with slope $0.0064\times 2N $ per $\pi/2$ corner of subregion $A$ at large $N$~\cite{casiniUniversal2006}. The red line in Fig.~\ref{fig:fig4} illustrates this large-$N$ expectation for four $\pi/2$ corners, the slope of which agrees with our data at all $N$ values investigated considering numerical uncertainties.}
\lj{Moreover, the fit of $4s(\pi/2)/N$ as a function of $1/N$, shown in the inset of Fig.~\ref{fig:fig4}, gives $4s(\pi/2)/N = 0.050(2) - 0.17(2)/N + O(1/N^2)$, in agreement with the large-$N$ expectation $\lim_{N \to \infty} 4s(\pi/2)/N = 0.0512$.}

The consistency of $s(\pi)$ with a CFT for $N \geq 8$, together with the agreement of the value of $s(\pi/2)$ for \lj{large $N$} with the field-theory expectation, serve as evidence that the transition in the SU($N$) lattice model for $N \geq \Nc$ realizes a genuine DQCP, described by the $N$-component Abelian-Higgs field theory.

\noindent{{\textcolor{blue}{\it Discussion.}}---}%
We have numerically studied the scaling of EE in a series of SU($N$) spin models, realizing direct transitions between SU$(N)$ N\'eel and VBS phases. By analyzing the subleading logarithmic corrections, we find that for relatively small values of $N$ (including $N=2,3,5,7$) the transition can not be described by a CFT, while for \mhs{larger values $N=8,10,...,18,20$} the EE scaling is compatible with a CFT description. These observations suggest the existence of a critical value $\Nc$ above which the SU$(N)$ DQCP is realized as a true continuous transition. Taking $N$ as a continuous variable, our data suggest that $\Nc$ lies between $7$ and $8$.
%
%$8$ and $10$.

A recent preprint~\cite{emidioEntanglement2024} studied the second-order R\'enyi entropy of the SU(2) $J$-$Q$ model with a different smooth cut from our current work, i.e. instead of a straight smooth cut that is along either the $\hat{x}$ or $\hat{y}$ direction, the reference made a ``tilted'' smooth cut that is along the $\hat{x}+\hat{y}$ or $\hat{x} - \hat{y}$ direction. Within error bar it was found that the logarithmic correction to the perimeter law vanishes in this case. However, in an upcoming work we will show that even with a tilted smooth cut, the subleading logarithmic contribution still exists in the 3rd and 4th Renyi entropy of the SU(2) $J$-$Q$ model, though the coefficient $s$ is smaller than the one for the straight smooth cut. Hence we expect that the existence of logarithmic subleading contribution to EE with smooth boundary is indeed ubiquitous for $N < \Nc$, regardless of the direction of the cut. But the direction dependence of the coefficient of the logarithmic term remains as a puzzle which needs to be addressed in future studies.

A candidate field theory for the family of SU$(N)$ DQCPs is the $N$-component Abelian-Higgs model. Four-loop renormalization group calculations~\cite{ihrigAbelian2019} suggest that the theory has a stable and real fixed point for $N\geq \Nc=12(4)$, which then collides  with a bicritical fixed point for $N \searrow \Nc$. This $\Nc$ is also compatible with numerical results for a lattice version of the Abelian-Higgs model~\cite{bonatiLattice2021}. The value is in fact close to the estimated $\Nc$ from the EE measurements. For $N < \Nc$, the two fixed points annihilate and disappear into the complex plane, leaving behind a weakly-first-order transition governed by 
%Miransky scaling
``walking behavior''~\cite{gorbenkoWalkingI2018}. 
This is illustrated in Fig.~\ref{fig:fig5}, which shows the schematic renormalization group flow of the Abelian-Higgs model for different values of $N$. 
Here, the renormalization group coupling $\lambda$ can be understood to parametrize the quartic self-interaction of the complex order-parameter field.
%
%The absence of a stable fixed point in the real coupling space for $N < \Nc$ is consistent with the observation of a negative log-coefficient $s(\pi)<0$ for small $N=2,3,5,7$. 
The ``walking behavior'' for $N < \Nc$ is one possible explanation of the observed anomalous logarithmic subleading terms of EE with smooth boundary~\footnote{The authors thank T.\ Senthil and Max Metlitski for proposing this possible explanation.}. 
%However, it is unknown whether the Abelian Higgs theory can actually explain the log-correction observed here at a more quantitative level.

\begin{figure}[tb!]
\includegraphics[width=\columnwidth]{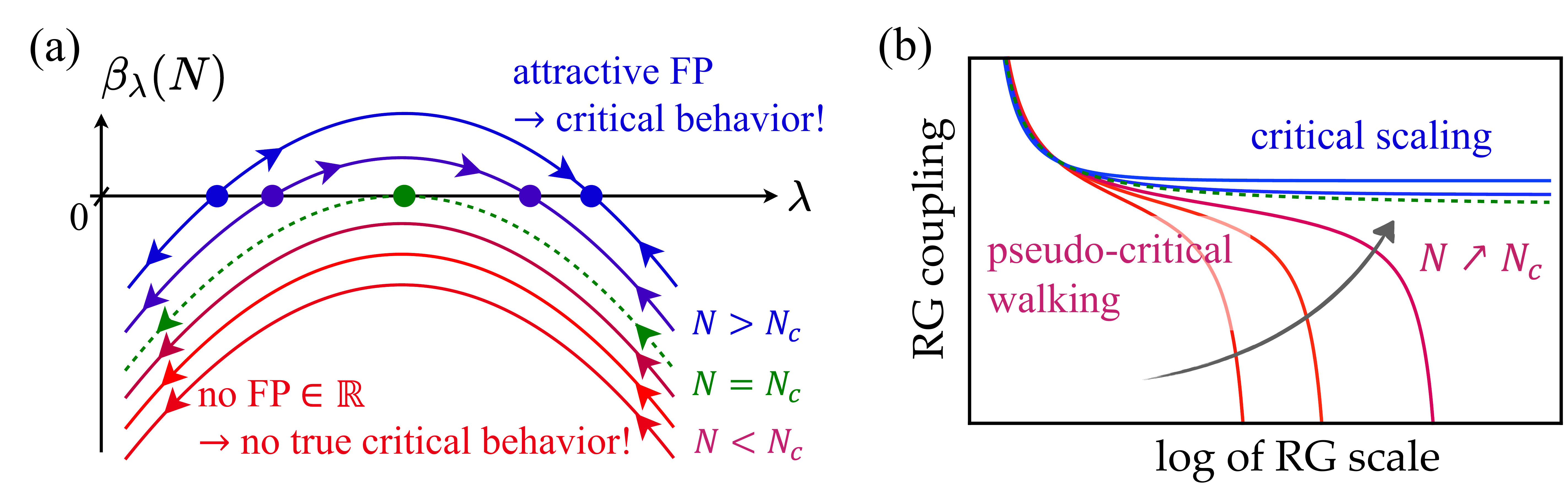}
\caption{\textbf{Illustration of fixed-point collision scenario.} 
(a)~Schematic renormalization group $\beta$ functions for the coupling $\lambda$, representing universal field theories that effectively describe the deconfined quantum phase transition for different values of $N$.
An example would be the quartic scalar coupling of the $N$-component Abelian-Higgs model.
%, see the discussion in the main text.
Corresponding renormalization group flow trajectories are shown in~(b). 
For $N>\Nc$, there are two fixed points, shown as blue dots in~(a). 
The attractive one leads to true critical scaling, as indicated by the blue renormalization group trajectories in~(b). 
Decreasing $N$ shifts the $\beta$ function down, until the two fixed points collide at some critical $N=\Nc$, indicated by the green dashed curves in (a) and (b).
Decreasing $N$ further, the fixed points annihilate and disappear into the complex plane, i.e., no true critical behavior can occur anymore. However, for $N\lesssim \Nc$, the renormalization group flow remains slow in the vicinity of the now complex fixed points, giving rise to walking behavior and drifting in the exponents, see red curves in (a) and (b).
}
\label{fig:fig5}
\end{figure}

Another possible mechanism for the observed $s(\pi)<0$ is the Goldstone modes from spontaneous symmetry breaking. 
%It was shown that in a SO$(n)$
In a SSB state with $n_\text{G}$ Goldstone modes, the EE indeed has a subleading logarithmic correction with coefficient 
%$s=-\frac{n-1}{2}$ 
$s_\text{G}=-n_\text{G}/2$ 
%independent of the geometry of the entanglement
when the subregion has a smooth boundary~\cite{metlitski2015entanglement}. For the SU(2) DQCP in the $J$-$Q_3$ model, this scenario was recently investigated in Ref.~\cite{deng2024diagnosing}. By including finite-size corrections in the formula for the scaling of EE in the SSB phase, it was found that the anomalous EE scaling may be captured by a weak SSB of the emergent SO(5) symmetry with four Goldstone modes, giving $s_\text{G}=-2$. \mhs{It is also interesting to notice that our $\scee$ data is non-linear against either $\ln L$ or $1/L$ for $N<N_c$ (see Fig.~\ref{fig:sm10} in SM).  In fact, for small $N$, we observe that $\scee$ scales linearly with $L$ (see Fig.~\ref{fig:sm11} in SM), implying that the perimeter law coefficients for smooth and corner cuts are different even though both cuts possess the same boundary length $l_A=2L$. A possible source for this anomaly may be the unequal critical fluctuations of the remaining VBS moments in different cuttings~\cite{emidioEntanglement2024}, which also points to a weekly first-order scenario at $N<N_c$. Whether the above phenomena persist at other DQCPs and bipartitions is worth studying in future works.}

Lastly, there is a distinct possibility that the logarithmic correction originates from near-marginal renormalization group flow on the entanglement cut. More precisely, the R\'enyi entropy can be viewed as the expectation value of a R\'enyi defect operator. The CFT result $s(\pi)=0$ holds in the deep IR limit of both the bulk and the defect. In other words, it assumes that the defect is conformal. However, for finite-size calculations and when the renormalization group flow on the defect is governed by nearly-marginal operators, for a window of system sizes logarithmic behavior can arise. However, it is difficult to explain in this scenario why anomalous logarithmic corrections are observed for several values of $N=2,3,5,7$, as the scaling dimensions of operators on defects should change with $N$, and unlikely to remain nearly marginal for these different values of $N$. Thus we conclude that the anomalous corrections are unlikely due to defect renormalization group flows, and should be attributed to bulk properties.

%More recently, a conformal 2D SU(2) DQCP with an SO(5)$_\text{f} \times$ SO(5)$_\text{b}$ global symmetry has been proposed as a description of the cuprate phase diagram with pseudogap metallic, $d$-wave superconducting, and charge ordered states as symmetry-breaking phases~\cite{christosModel2023}. Investigating the validity of this proposal using entanglement measurements, as outlined in this work, represents an interesting direction for future research. \meng{(ZYM:I discuss the work from Subir in the early version, not sure whether we still need it.)}

\noindent{\textcolor{blue}{\it{Acknowledgments.---}}}%
We thank Fakher Assaad, Jonathan D'Emidio, Yin-Chen He, Max Metlitski, Subir Sachdev, Anders Sandvik, and Kai Sun for valuable discussions on related topics. 
{We are grateful to Max Metlitski for pointing out the possibility of defect renormalization group flows, and also for discussions concerning possible mechanisms of log-corrections in the pseudocriticality scenario.}
MHS, JRZ and ZYM thank Jonathan D'Emidio, Ting-Tung Wang and Yuan Da Liao for fruitful discussions on algorithm development and implementation. They acknowledge the support from the
Research Grants Council of Hong Kong (Project Nos. AoE/P701/20, 17309822,
C7037-22GF, 17302223, 17301924), the
ANR/RGC Joint Research Scheme sponsored by Research
Grants Council of Hong Kong and French
National Research Agency (Project No. A\_HKU703/22), the GD-NSF (No.\ 2022A1515011007) and the HKU Seed Funding for Strategic Interdisciplinary
Research.
The work of LJ is supported by the Deutsche Forschungsgemeinschaft (DFG) through SFB 1143 (A07, Project No.\ 247310070), the W\"urzburg-Dresden Cluster of Excellence \emph{ct.qmat} (EXC 2147, Project No.\ 390858490), and the Emmy Noether program (JA2306/4-1, Project No.\ 411750675).
MMS acknowledges support from the Deutsche Forschungsgemeinschaft (DFG) through SFB 1238 (C02, Project No.\ 277146847) and the DFG Heisenberg program (Project No.\ 452976698). C.X. is supported by the Simons foundation through the Simons Investigator program. M.C. acknowledges supports from NSF under award number DMR-1846109. 
The authors also acknowledge the Tianhe-II platform at the National Supercomputer Center in Guangzhou, the HPC2021 system under the Information Technology Services, University of Hong Kong, the Beijng PARATERA Tech CO., Ltd.\ (URL: https://cloud.paratera.com), and the Center for Information Services and High Performance Computing (ZIH) at TU Dresden, which is jointly supported by the German Federal Ministry of Education and Research and the state governments participating in the NHR (URL: https://www.nhr-verein.de/unsere-partner) for providing HPC
resources that have contributed to the results reported in this paper.

\bibliographystyle{longapsrev4-2}
\bibliography{bibtex}

%%%%%%%%%%%%%%%%%%%%%%%%%%%%%%%%%%%%%%%%%%%%%%%%%%%%%%%%%%%%%%%%%%%%%%%%%%
% SUPPLEMENTARY MATERIALS
%%%%%%%%%%%%%%%%%%%%%%%%%%%%%%%%%%%%%%%%%%%%%%%%%%%%%%%%%%%%%%%%%%%%%%%%%%
\clearpage
\onecolumngrid

%\appendix
\setcounter{equation}{0}
\setcounter{figure}{0}
\setcounter{table}{0}
\setcounter{page}{1}
\makeatletter
\renewcommand{\theequation}{S\arabic{equation}}
\renewcommand{\thefigure}{S\arabic{figure}}
\setcounter{secnumdepth}{3}	

\begin{center}
\bf \uppercase{Supplementary Materials for ``Deconfined quantum criticality lost''}
\end{center}
\vspace{2\baselineskip}

\twocolumngrid

\section{QMC implementation}
QMC simulations for the SU($N$) spin models are generalizations of the SU(2) cases~\cite{kaulLattice2012,BeachcontinuousN2009,Harada2003,Kaul2011,haradaPossibility2013,jonNoneq2020,jonspinchain2015}. In particular, there are $N$ colors for spins and loops in the loop algorithm~\cite{sandvikLoop2010}. 
Since all the off-diagonal elements in the Hamiltonian in Eq.~\eqref{eq:eq1} in the main text are negative, the model can be simulated without a sign problem if off-diagonal operators appear an even number of times along the imaginary time direction, as in the SU(2) case. 

Note that in Eq.~\eqref{eq:eq1}, the projection operator $P_{ij}$ [the SU($N$) generalization of the nearest-neighbor antiferromagnetic interaction] only acts between spins belonging to different sublattices. In contrast, $\Pi_{ij}$ [the SU($N$) generalization of the the next-nearest-neighbor ferromagnetic interaction] only acts between spins belonging to the same sublattice. One can decompose $P_{ij}$ and $\Pi_{ij}$ into diagonal and off-diagonal parts, namely $P_{ij}= P_{ij}^1-P_{ij}^2$ and $\Pi_{ij}= \Pi_{ij}^1-\Pi_{ij}^2$, where $1$ and $2$ labels diagonal and off-diagonal parts, respectively. Therefore, all the non-zero matrix elements can be explicitly computed as 
$\left \langle \alpha_A\alpha_B| P^1_{ij}|\alpha_A\alpha_B  \right \rangle = \left \langle \alpha_A\alpha_A| \Pi^1_{ij}|\alpha_A\alpha_A  \right \rangle =\left \langle \beta_A\beta_B| P^2_{ij}|\alpha_A\alpha_B  \right \rangle =\left \langle \beta_A\alpha_A| \Pi^1_{ij}|\alpha_A\beta_A  \right \rangle = \frac{1}{N}$, where $|\alpha \rangle$, $|\beta\rangle$ denotes two out of $N$ possible colors of a spin, and the subscripts $A$ and $B$ denote the different sublattices. In particular, the diagonal operators $P^1_{ij}$ and $\Pi^1_{ij}$ act only between spins with the same color and leave the state intact. $P^2_{ij}$ simultaneously changes the color of two spins with the same color, $\left | \alpha_A\alpha_B  \right \rangle \to\left | \beta_A\beta_B  \right \rangle$, and $\Pi^2_{ij}$ permutes the colors of two spins with different colors, $\left | \alpha_A\beta_A  \right \rangle \to\left | \beta_A\alpha_A  \right \rangle$.

Let us outline the stochastic series expansion (SSE) QMC sampling process and introduce the generalized loop update for the SU($N$) cases. At the start of each Monte Carlo step, one performs a diagonal update where $P_{ij}^1$ or $\Pi_{ij}^1$ is inserted or removed with Metropolis probability determined by the matrix elements listed above. Next, linked vertices are constructed to form loops in the configuration. Then, a random color and a starting position are picked. One follows the trajectory of the loop and paints the visited spins with the loop color until the loop closes. An exchange between the diagonal and off-diagonal operators may happen during the painting. Once a loop closes, a new configuration is generated, and one can always accept the update since all the non-zero matrix elements are equal to $1/N$ and thus share the same weight. Finally, one performs measurements within the new configuration. 

\begin{figure}[tb!]
	\includegraphics[width=\columnwidth]{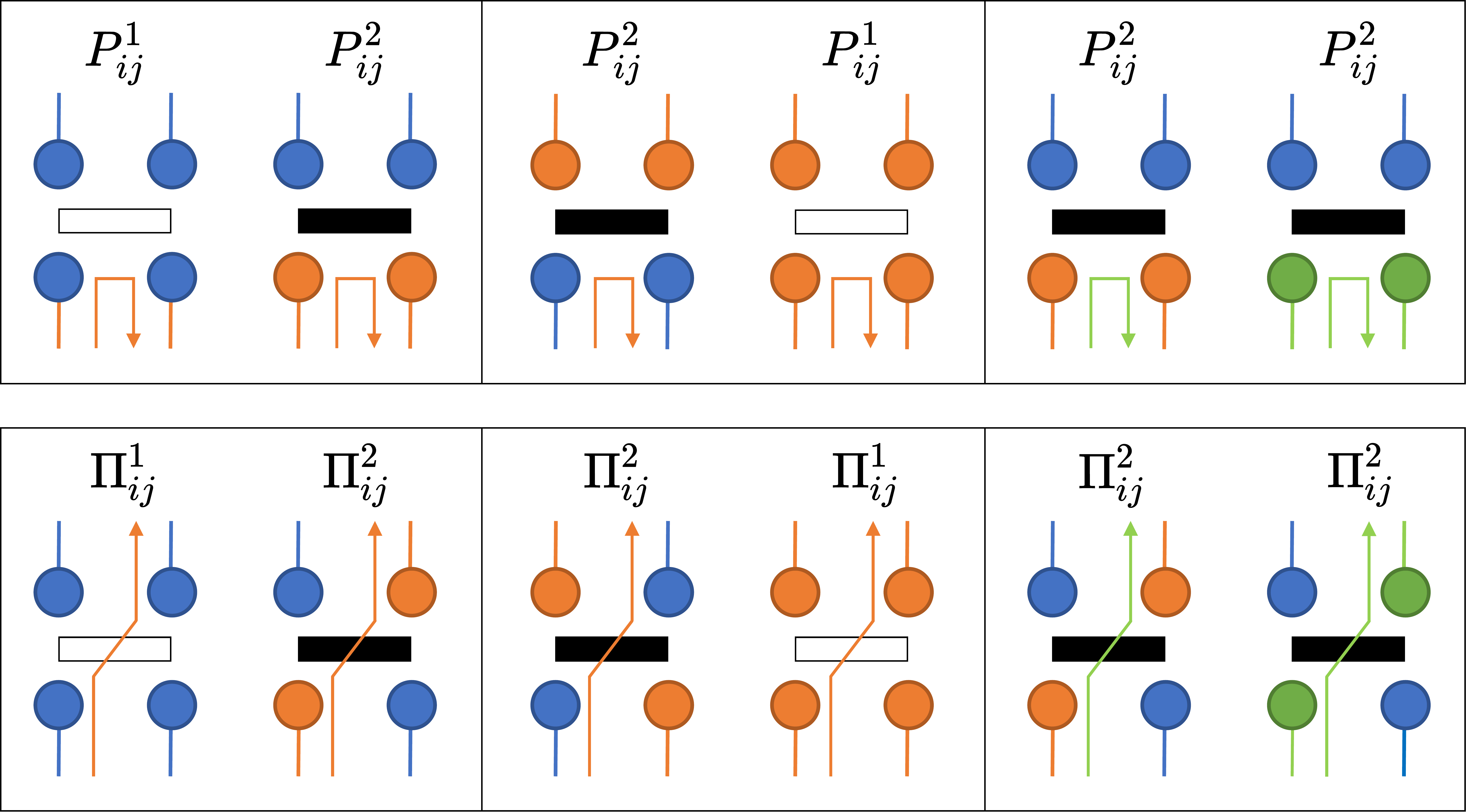}
	\caption{\textbf{Typical vertices of the SU($\boldsymbol{N}$) Hamiltonian and different loop updating moves.} The left side of each block is the vertex before the update, and the right side shows the vertex after the update. The upper panel demonstrates the vertices involving nearest-neighbor spin interaction $P_{ij}$ (switch and reversed move). The lower panel exhibits those for the next-nearest-neighbor interaction $\Pi_{ij}$ (switch and continued move). Arrows with colors represent the trajectories of colored loops, which paint the spins. Diagonal operator (white bar, $P^{1}_{ij}$ and $\Pi^{1}_{ij}$) and off-diagonal operator (black bar, $P^{2}_{ij}$ and $\Pi^{2}_{ij}$) may transform each other to ensure a non-trivial configuration after the update.}
	\label{fig:sm1}
\end{figure}

Loop moves are designed to avoid zero-weighted configurations for high sampling efficiency. Figure~\ref{fig:sm1} shows typical vertices that may occur in simulating Hamiltonians with $P_{ij}$ and $\Pi_{ij}$. The upper panel shows the vertices with $P_{ij}$ requiring a switch and reversed loop move. The lower one shows the vertices with $\Pi_{ij}$ requiring a switch and continued loop move. The left vertex in each block is before the update, along with the path and color of the loop represented by the colored arrows. The right side of each block illustrates the vertex after the update. The spins visited by the loop are painted with the color of the loop. Meanwhile, the type of the operator is considered to be changed to ensure the new configuration is non-trivial.

\begin{figure}[tb!]
	\includegraphics[width=\columnwidth]{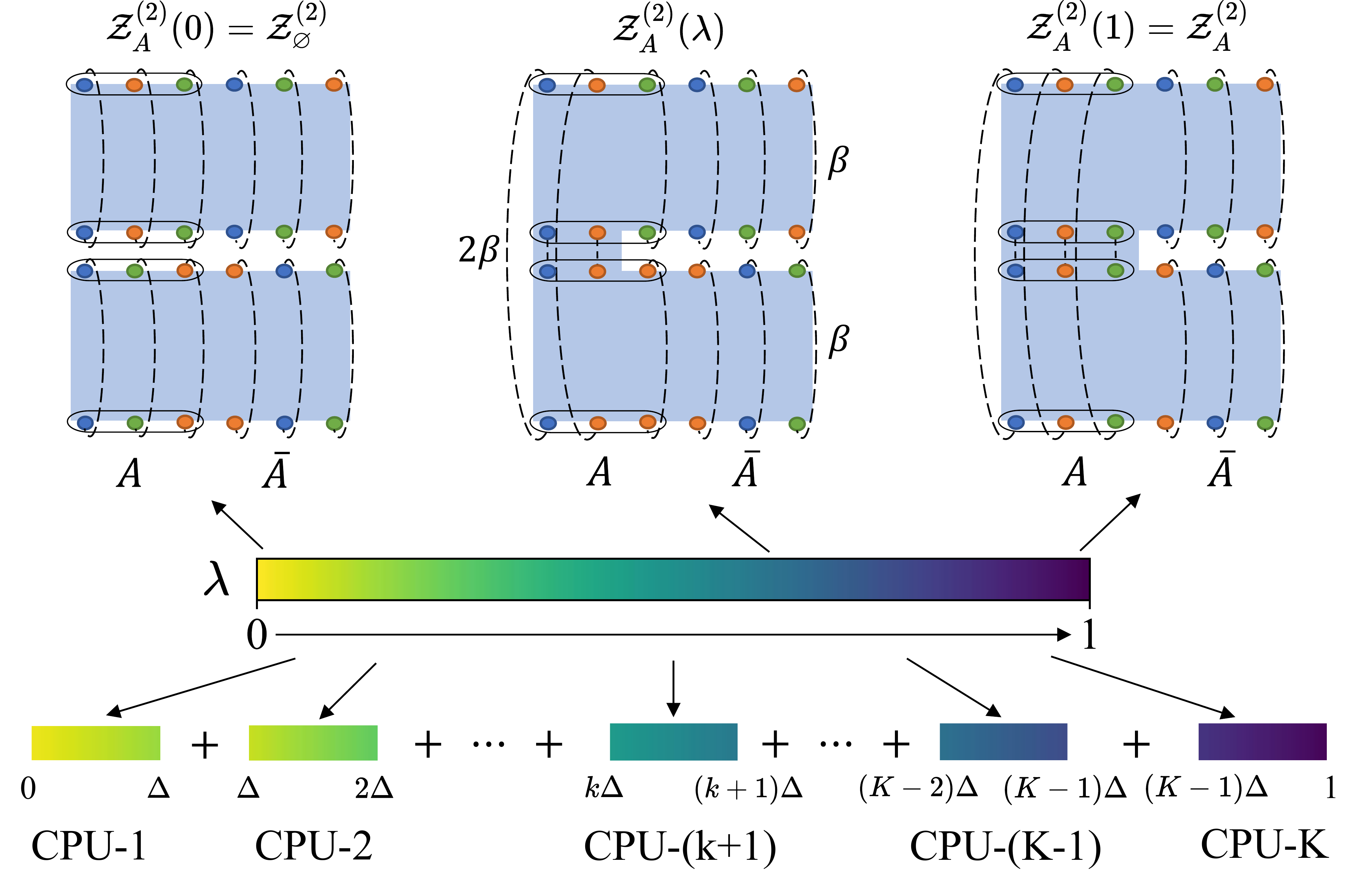}
	\caption{
		\textbf{Incremental computation of EE.}  As in Eq.~\eqref{eq:eqS1}, we split the computation of the ratio of partition functions into the parallel execution of many ratios and compute the EE via a non-equilibrium process characterized by $\lambda$ evolving from $\lambda=0$ to $\lambda=1$. We divide this path into $K$ pieces and assign each piece to one CPU. The connectivity of spins in the entanglement region $A$, depicted as black circles, between two replicas (or topology of the joint partition function) is determined stochastically by $\lambda$. As $\lambda$ increases, more spins in $A$ from different replicas will be `glued' together, given that they share the same color, resulting in an imaginary time period of $2\beta$. Spins not `glued' together are in the environment $\overline{A}$ and experience a regular imaginary time period of $\beta$.}
	\label{fig:sm2}
\end{figure}

\section{Algorithms for entanglement entropy of SU($N$) spins}
\label{algorithms}
We implement the recently developed incremental algorithm to compute the EE~\cite{jonNoneq2020,zhaoScaling2022,zhaoMeasuring2022,emidioUniversal2022,liaoTeaching2023,panComputing2023,songQuantum2023} and generalize it to SU($N$) spin models. We first parameterize the partition function $\mathcal{Z}_{A}^{(2)}$ with $\lambda$ such that $\mathcal{Z}_{A}^{(2)}(\lambda=0)=\mathcal{Z}_\emptyset^{(2)}$ and $\mathcal{Z}_{A}^{(2)}(\lambda=1)=\mathcal{Z}_A^{(2)}$. More explicitly, $\mathcal{Z}_{A}^{(n)}(\lambda)=\sum_{B \subseteq A}g_{A}\left(\lambda,N_{B}\right) \mathcal Z_{B}^{(n)}$ where $B$ is a subset of the entanglement region $A$, $N_B$ is the number of sites in $B$ and $g_A(\lambda,N_B)=\lambda^{N_B}(1-\lambda)^{N_A-N_B}$ with $\lambda \in[0,1]$. Therefore, $S^{(n)}_A$ can be written as the integral $S_{A}^{(n)}=\frac{1}{1-n} \int_{0}^{1} d \lambda \frac{\partial \ln \mathcal{Z}_{A}^{(n)}(\lambda)}{\partial \lambda}$ along the path $\lambda \in[0,1]$. Instead of calculating $e^{-S^{(2)}_A}$ directly, we further split this path into $K$ pieces with a step length $\Delta$, the ratio of partition function can now be written as
\begin{equation}
	\begin{aligned}
	e^{-S_A^{(2)}}=\frac{\mathcal{Z}(1)}{\mathcal{Z}(0)}=&\frac{\mathcal{Z}(\Delta)}{\mathcal{\mathcal{Z}}(0)}\frac{\mathcal{Z}(2\Delta)}{\mathcal{Z}(\Delta)}\\&\cdots\frac{\mathcal{Z}(k\Delta)}{\mathcal{Z}((k-1)\Delta)} \cdots\frac{\mathcal{Z}(1)}{\mathcal{Z}((K-1)\Delta)},
	\end{aligned}
\label{eq:eqS1}
\end{equation}
where we have suppressed the R\'enyi index in the intermediate $\mathcal{Z}$'s on the right-hand side of Eq.~\eqref{eq:eqS1} for simplicity. Each term in the product string, with a well-controlled value of $O(1)$ instead of exponentially small in the left-hand side of Eq.~\eqref{eq:eqS1}, is computed in parallel, as shown in Fig.~\ref{fig:sm2}. Finally, we multiply these pieces and take the logarithm to get the R\'enyi entropy $S^{(2)}_A$.

To implement the algorithm, we first thermalize the regular partition function $\mathcal{Z}$ and then make two replicas of it as the thermalized configuration of $\mathcal{Z}_{\varnothing}^{(2)}$ (the leftmost configuration in Fig.~\ref{fig:sm2}). We divide the interval $\lambda \in[0,1]$ into $K$ pieces with a length of each sub-interval $\Delta$ and distribute each process to one CPU, as shown in Fig.~\ref{fig:sm2}. Take process $k+1$ as an example,  $\lambda$ evolves from $\lambda(t_{\mathrm i})=k\Delta$ to $\lambda(t_{\mathrm f})=(k+1)\Delta$. At each $\lambda$ value, we need to determine the topology of $\mathcal{Z}_{A}^{(2)}(\lambda)$. Each site in $A$ is considered to be `glued' or `separated' according to the probability  $P_{\text{join}}=\min \left\{\frac{\lambda}{1-\lambda}, 1\right\}$ and $P_{\text{leave}}=\min \left\{\frac{1-\lambda}{\lambda}, 1\right\}$, with the condition that spins from two replicas at that site share an identical color. After determining the trace structure, we perform a Monte Carlo update and take measurements. To take non-equilibrium measurements, we gradually increase $\lambda(t_{\mathrm i})=k\Delta$ by a small value $h$ and record the value $\frac{g_{A}\left(\lambda\left(t_{m+1}\right), N_{B}\left(t_{m}\right)\right)}{g_{A}\left(\lambda\left(t_{m}\right), N_{B}\left(t_{m}\right)\right)}$, where $\lambda(t_m)=k\Delta+mh$. Each time we increase $\lambda$, the topology of $\mathcal{Z}_{A}^{(2)}(\lambda)$ should be re-determined. The process is repeated until $\lambda(t_m)=\lambda(t_{\mathrm f})=(k+1)\Delta$. At the end of process $k+1$, we compute $\frac{\mathcal{Z}((k+1)\Delta)}{\mathcal{Z}(k\Delta)}=\left\langle\prod_{m=0}^{\Delta /h-1} \frac{g_{A}\left(\lambda\left(t_{m+1}\right), N_{B}\left(t_{m}\right)\right)}{g_{A}\left(\lambda\left(t_{m}\right), N_{B}\left(t_{m}\right)\right)}\right\rangle .$ Finally, we multiply the results all processes together to obtain $S^{(2)}_A$ using Eq.~\eqref{eq:eqS1}. 

It is important to note that, although $e^{-S^{(2)}_A}(L)$ is an exponentially small number and therefore exponentially hard to be sampled well as the system size increases, each term on the right-hand side of Eq.~\eqref{eq:eqS1} is of $O(1)$ and therefore easy to compute precisely~\cite{panComputing2023}. Their product can then be computed accurately, and one then takes its logarithm to obtain the $S^{(2)}_A(l)$. The divide-and-conquer strategy of the incremental algorithm guarantees the precise determination of the $S^{(2)}_A(l)$, such that one can analyze its finite-size scaling behavior and find the log-coefficient in Eq.~\eqref{eq:eq2} in the main text.

\mhs{In the main text, we also employ the recently developed subtracted corner entanglement entropy, $\scee$ to obtain the subleading log-coefficient at large values of $N$, in one Monte Carlo simulation~\cite{liao2024extracting}. The $\scee$ is defined as $\scee = S^{(2)}_\mathrm{smooth}-S^{(2)}_\mathrm{corner}$ where $S^{(2)}_{\mathrm{smooth}}$ and $S^{(2)}_{\mathrm{corner}}$ are the second order R\'enyi entropy for the red and blue region in the inset of Fig.~\ref{fig:fig3}(b) in the main text. In principle, the difference of EEs between two subregions, $A_1$ and $A_2\supset A_1$ can be calculated in a similar manner using non-equilibrium method. One can write $\ssr_{A_2} - \ssr_{A_1} = -\ln \frac{\zr_{A_2}}{\zr_{\emptyset}}+\ln \frac{\zr_{A_1}}{\zr_{\emptyset}} = -\ln \frac{\zr_{A_2}}{\zr_{A_1}}$. The ratio $\zr_{A_2}/\zr_{A_1}$ is regarded as the work done when quenching slowly from $\zr_{A_1}$ to $\zr_{A_2}$. Now, one can define a $\lambda$-parametrized partition function $$\zr_{A_2-A_1}(\lambda) = \sum_{B\subseteq A_2\setminus A_1} \lambda^{N_B}(1-\lambda)^{N_{A_1}-N_{A_2}-N_{B}}\zr_{B+A_1}$$ with $\zr_{A_2-A_1}(0) =  \zr_{A_1}$ and $\zr_{A_2-A_1}(1) = \zr_{A_2}$. $\zr_{B+A_1}$ is the partition function with region $B$ and $A_1$ glued together in imaginary time. Hence, one can compute incrementally $\ssr_{A_2} - \ssr_{A_1} = -\int_0^{1}d\lambda \frac{\partial \ln \zr_{A_2-A_1}(\lambda)}{\partial\lambda}$ using the same routine discussed above.} 

{In this way, we don't need to compute the $\ssr_{A_2}$ and $\ssr_{A_1}$ individually and take their difference, but can cancel out the perimeter law contribution during the Monte Carlo simulation and hence save the computational efforts and resources. Such a approach of $\scee$ has been proved to be successful and offered the most accurate estimation of the universal corner entanglement entropy coefficient for the (2+1)D O(3) QCP~\cite{liao2024extracting}.}

\begin{figure}[tbp!]
	\includegraphics[width=0.95\columnwidth]{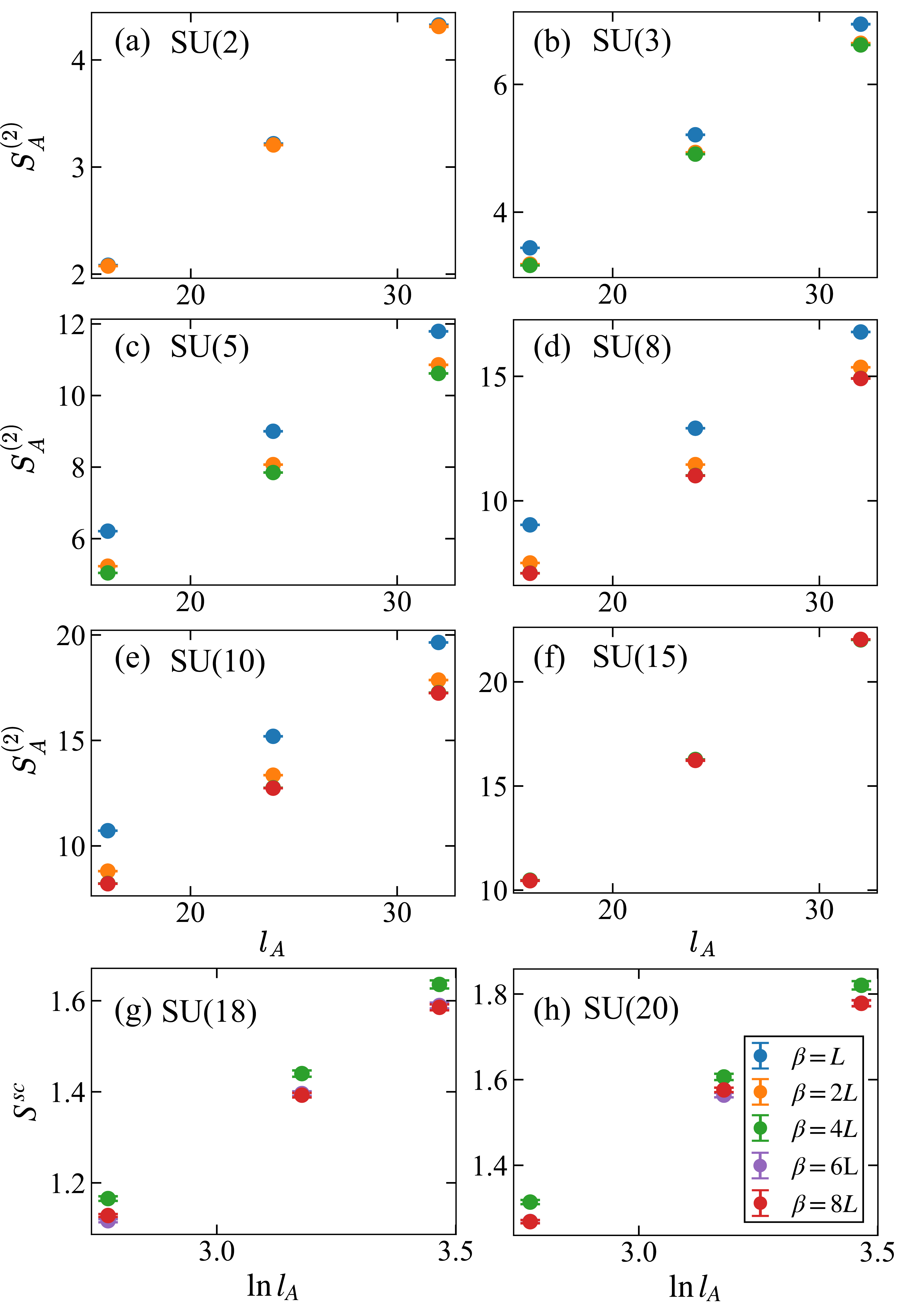}
	\caption{
		\textbf{Convergence of EE and $\scee$ against temperature for various $N$.} \mhs{(a)-(f) demonstrates how EE changes against temperature for various $N$. (g)-(h) presents the convergence of $\scee$ against temeprature. The system sizes presented are $L=8,12,16$. All panels share the legend in (h).}}
	\label{fig:T_converge}
\end{figure}  

To ensure we are computing the EE at a sufficiently low temperature without thermal pollution, we investigate the convergence of EE by keeping $\beta=nL$ with different integer $n$s. Fig.~\ref{fig:T_converge} demonstrates the convergence of EE at various $N$. For the SU(2) case, EE is converged using $\beta=L$. As $N$ increases, $\beta=L$ is no longer a sufficient low temperature to compute EE. We notice that the larger the $N$ is, the lower the temperature one should keep to ensure convergence. \mhs{For example, at $N=15$, we show that keeping $\beta=4L$ and $\beta=8L$ obtain the same EE value within error bars (two sets of data overlap in panel (f)). Therefore, keeping $\beta=4L$ is sufficient for $N\le 15$. In our simulation, we keep $\beta=L$ for the SU(2) case and $\beta=4L$ for other $N\le 15$. At $N=18,20$, where we compute $\scee$, $\beta=4L$ is no longer sufficient as shown in Fig.~\ref{fig:T_converge}(g)(h) and we keep $\beta=8L$ in computing $\scee$.}

\begin{figure*}[tbp!]
	\includegraphics[width=0.7\textwidth]{sm3}
	\caption{
		\textbf{Stochastic data collapse to determine critical exponents.} Color plots display distribution of fitting error $\delta$ in $\nu$-$\beta$ plane for SU(3), SU(5) and SU(15) from left to right, using different maximal lattice sizes $L_\text{max} = 48$, $84$, and $108$ from top to bottom. Black dot in each panel indicates the optimal set of exponents in each case.}
	\label{fig:sm3}
\end{figure*}  

\begin{figure*}[tbp!]
	\includegraphics[width=0.7\textwidth]{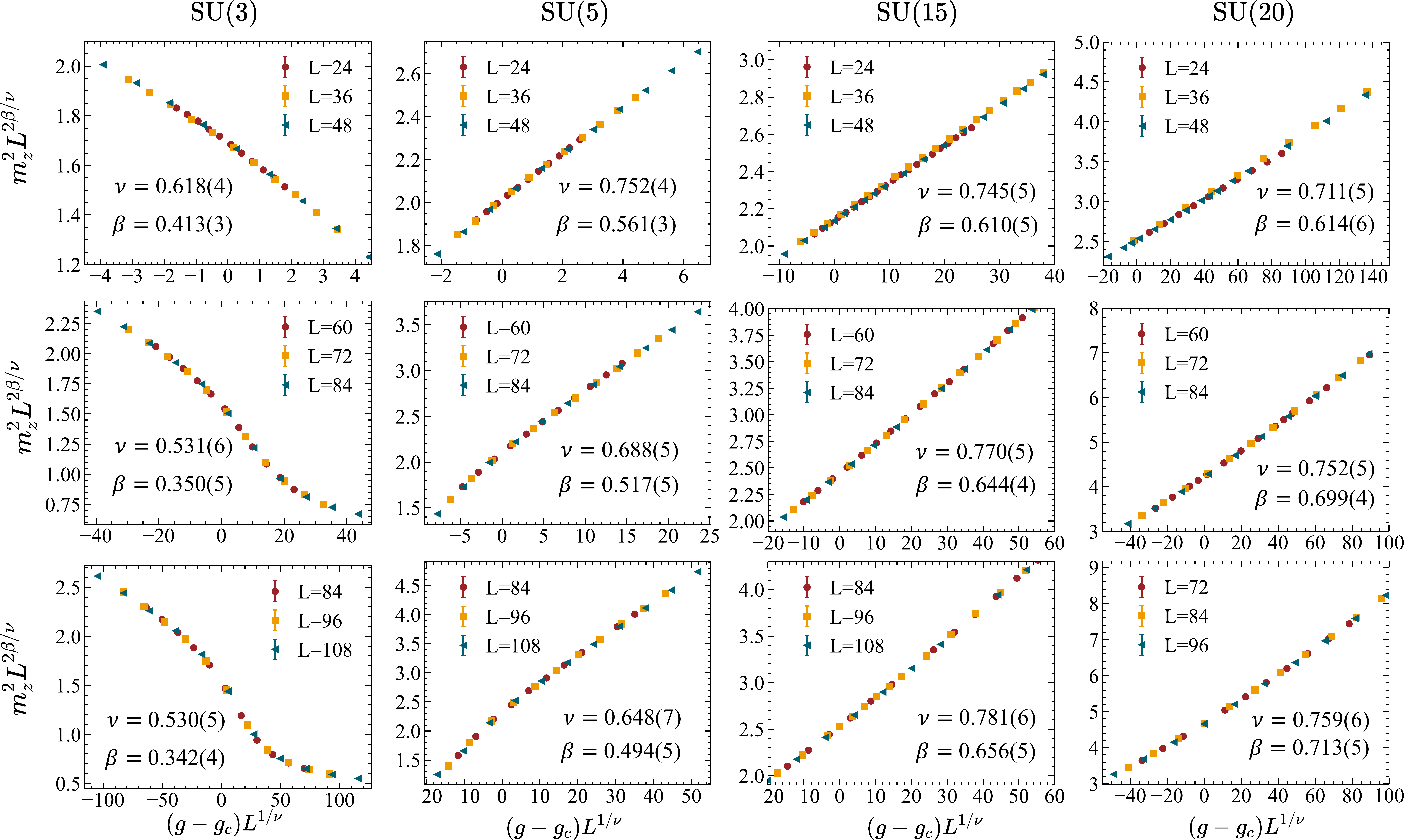}
	\caption{
		\textbf{Data collapse of squared magnetization.} Panels show data collapses of squared magnetization $m_z^2$ for SU(3), SU(5) and SU(15) from left to right, from three consecutive system sizes $L \in [L_\text{max} - 24, L_\text{max}]$ with $L_\text{max} = 48$, $84$, and $108$ from top to bottom, with the optimal $\beta$ and $\nu$ as determined in Fig.~\ref{fig:sm3}.}
	\label{fig:sm4}
\end{figure*}  

\section{Stochastic data collapse}
\label{stochastic}
We have devised a method for accurately estimating critical exponents, which involves collapsing data using a stochastic process~\cite{wangEmus2023,yanFully2022,songQuantum2023}. This involves fitting a polynomial curve through the data points for various system sizes $L$, and the quality of the collapse is determined by how well the data fits the curve. To quantify this, we use the R-squared value, denoted by $R^2$, representing the variation between the data and the fitted curve. Its definition is $R^2 = 1-\frac{S_\mathrm{res}}{S_\mathrm{tot}}=1-\delta$, with $S_\mathrm{res}=\sum_{i=1}^{n}w_{i}\left(y_{i}-\hat{y}_{i}\right)^{2}$ and $S_\mathrm{tot}=\sum_{i=1}^{n}w_{i}\left(y_{i}-\bar{y}\right)^{2}$. The smaller the value of $\delta$, the smaller the error of the fitting and the better the quality of the collapse. $S_\mathrm{res}$ measures the deviation between the actual data and the fitted curve, whereas $S_\mathrm{tot}$ measures the variance of the fitted curve itself. The weight $w_i$ is used to emphasize the importance of the critical region, where the quality of the collapse is of utmost importance. The $y$ value of the scaled data point is denoted by $\hat{y}_i$, and that of the fitted curve at the same $x$ value is denoted by $y_i$. The fitted curve's mean value of all points $y_i$ is denoted by $\bar{y}$.

To investigate the drift of exponents against system sizes, we fix the critical point at the extrapolated value at $L\to \infty$ and use three different sizes $L_\mathrm{max}-24$, $L_\mathrm{max}-12$, and $L_\mathrm{max}$ together at a time to obtain the exponents. Then, we can set $\beta$ and $\nu$ as free parameters, and the stochastic process is done in the two-dimensional plane spanned by $\beta$ and $\nu$. A random set of parameters is proposed and fitted by a polynomial curve. The fitting error $\delta$ is calculated. Then one randomly moves parameters in the two-dimensional parameter space as shown in Fig.~\ref{fig:sm3}, while recording the fitting error $\delta$. After enough steps, the best estimate is the parameter set with the smallest error. Distributions of $\delta$ are exemplified in Fig.~\ref{fig:sm3} for SU(3), SU(5) and SU(15), for different $L_\mathrm{max}$. Figure~\ref{fig:sm4} illustrates the corresponding collapses of the squared magnetization $m_z^2$ using the fitted critical exponents $\nu$ and $\beta$. 

\begin{figure}[tbp!]
	\includegraphics[width=\columnwidth]{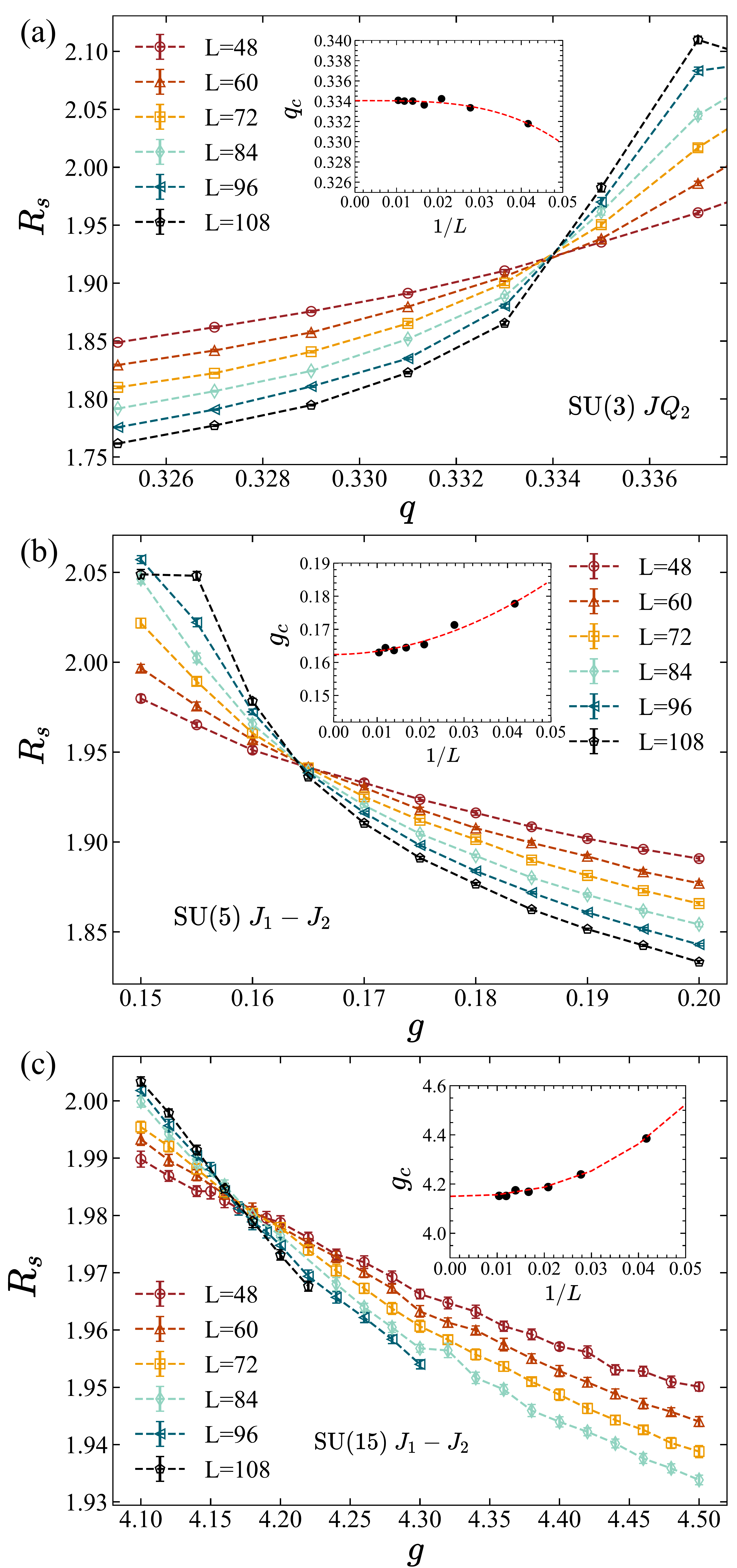}
	\caption{\textbf{Crossings of Binder ratios to determine the transition points $\boldsymbol{q_\mathrm{c}}$ and $\boldsymbol{g_\mathrm{c}}$ for SU(3), SU(5) and SU(15).} We use the Binder ratio for the antiferromagnetic N\'eel order to determine the critical $q=\frac{Q}{J_1+Q}$ for (a) the SU(3) case and the critical $g=\frac{J_2}{J_1}$ for (b)(c) the SU(5) and SU(15) cases, for different system sizes $L = 24, 34, 48, 60, 72, 84, 96, 108$ and inverse temperatures $\beta = L$. Insets show the crossing points $q_\mathrm c$ and $g_\mathrm c$ as function of $1/L$. The extrapolated values of the critical points in the thermodynamic limit are well consistent with previous work~\cite{louvbsneel2009,kaulLattice2012}.}
	\label{fig:sm5}
\end{figure} 

\section{QMC benchmark of DQCPs and drift of critical exponents}
In this section, we first show representative data in which the positions of the DQCPs are obtained from the crossing of the N\'eel order Binder ratios $R_{s}=\frac{\left\langle m_{ z}^{4}\right\rangle}{\left\langle m_{ z}^{2}\right\rangle^{2}}$ for different $N$. 
Figure~\ref{fig:sm5}(a) and (b) show, for $N=3$ and $5$, that the location of the transition points are consistent with previous works~\cite{kaulLattice2012,louvbsneel2009}. 
Figure~\ref{fig:sm5}(c) shows the corresponding data for $N=15$, for which no results were previously available in the literature. 
The EE computation discussed in the main text is performed at the so-determined transition points $q_\mathrm{c}$ and $g_\mathrm{c}$, respectively.

\begin{figure*}[tbp!]
\includegraphics[width=0.7\textwidth]{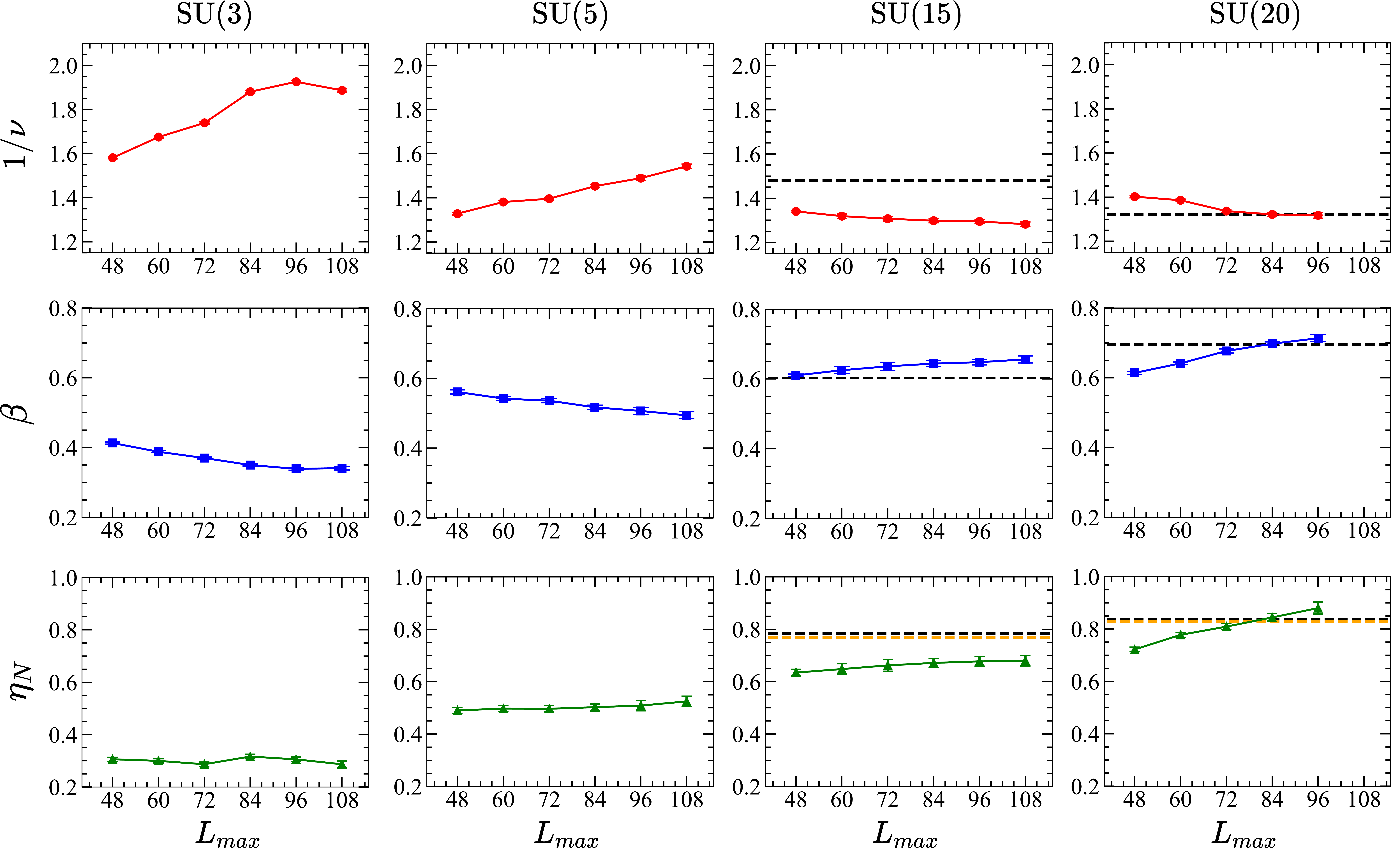}
\caption{\textbf{Drift of $1/\nu$, $\beta$ and $\eta_\mathrm{N}$ for SU(3), SU(5) and their convergence in SU(15) cases.} In the cases of the SU(3) and SU(5), one sees $1/\nu$ values are still not converged as the largest system size $L_\text{max}$ used in the stochastic data collapse analysis gradually increases, suggesting that these transitions are weakly-first-order. But in the case of SU(15), the exponents $1/\nu$ converge as $L_\text{max}$ increases. Green dashed lines are the large-$N$ prediction for the Abelian Higgs model: ${\nu}^{-1}=(1-\frac{48}{\pi^2 N})^{-1}$, $\eta_\mathrm{N}=1-\frac{32}{\pi^2 N}$ and $\beta=\nu(\eta_\mathrm{N}+1)/2$ at order $1/N$. Orange dashed lines show the large-$N$ prediction for $\eta_\text{N}$ at order $1/N^2$~\cite{kaulLattice2012}.}
	\label{fig:sm7}
\end{figure*}

Then, we further carry out the finite-size analysis based on stochastic data collapse~\cite{wangEmus2023,yanFully2022,songQuantum2023} as introduced in Sec.~\ref{stochastic} to determine the critical exponents as a function of the system size $L$, see Fig.~\ref{fig:sm7}. Previous work~\cite{haradaPossibility2013} has obtained critical exponents for small $N=2,3,4$ in the $J$-$Q$ model. They found a non-convergent and increasing trend of the exponent $1/\nu$ against system sizes, implying a weakly-first-order transition at $N=2,3,4$. Our results for $1/\nu$ for SU(3) and SU(5), as shown in Fig.~\ref{fig:sm7}, manifest a similar non-convergent and increasing behavior as in Ref.~\cite{haradaPossibility2013}. In fact, our estimated exponents for SU(3) match quantitatively well with those of Ref.~\cite{haradaPossibility2013}.  

On the other hand, when $N\ge 8$, where the DQCPs are consistent with CFTs as discussed in the main text, we find that the $1/\nu$ values converge quickly as a function of system sizes, showing no anomaly, see Fig.~\ref{fig:sm7} for SU(15) case. This observation strengthens our conclusion that SU$(N<\Nc)$ transitions are weakly first order, while SU$(N\ge \Nc)$ transitions, with $\Nc$ between $7$ and $8$, as obtained from the scaling of the EE for smooth boundaries, are continuous and well described by the Abelian Higgs or  noncompact CP$^{N-1}$ nonlinear sigma models.

The last row in Fig.~\ref{fig:sm7} shows our estimated $\eta_\text{N}=2\beta/\nu-1$ for the antiferromagnetic order. It is interesting to see that, in the SU(3) case, our $\eta \sim 0.3$ and in the SU(5) case, our $\eta \sim 0.5$, are all consistent with previous values~\cite{kaulQuantum2012} up to the studied system sizes. The dashed lines in Fig.~\ref{fig:sm7} indicate the large-$N$ prediction for the Abelian Higgs model: ${\nu}^{-1}=(1-\frac{48}{\pi^2 N})^{-1}$ and $\eta_\mathrm{N}=1-\frac{32}{\pi^2 N}$ at order $1/N$~\cite{halperinFirst1974, irkhinExpansion1996}. Our obtained $\eta_\text{N}$ values at $N=15$ follow the trend of the large-$N$ prediction, providing precious results in addition to those reported in Ref.~\cite{kaulLattice2012}.

\section{Fitting quality analysis}

In this section, we analyze the quality of the fitting of our data. To reliably remove the leading \lj{perimeter} law contribution and expose the subleading correction, we investigate both the subtracted EE~\cite{song2024extracting}, $S^{(2)}_A(2L)-2S^{(2)}_A(L)$ (\cx{whose perimeter law term is automatically canceled out}), and directly minus the leading \lj{perimeter} law contribution, $S^{(2)}_A-al_A$, using beforehand fitted coefficient $a$. We fit both quantities against $\ln L$ and $1/L$ to identify whether there is indeed an anomalous logarithmic subleading contribution, or a regular finite-size correction $\sim 1/L$. 
%subleading term is logarithmic or just arising from finite-size effects. 
%
{From Eq.~\eqref{eq:eq1} in the main text, we expect for a CFT, the EE scales as
\begin{align} \label{eq:eqsm2}
    S_{A}^{(2)}(L) - a L = - s \ln L + c' + O(1/L),
\end{align}
and 
\begin{align}\label{eq:eqsm3}
    S_{A}^{(2)}(2L) - 2S_A^{(2)}(L) = + s \ln L + c'' + O(1/L),
\end{align}
where $c'$ and $c''$ are nonuniversal constants.
As a consequence, the slopes of $S_{A}^{(2)}(2L) - 2S_A^{(2)}(L)$ and $S_{A}^{(2)}(L) - a L$ as function of $\ln L$ give the universal coefficient $s$ and $-s$, respectively, of the log-correction.}
The results for smooth boundaries are presented in Fig.~\ref{fig:sm8} for the subtracted EE $S_{A}^{(2)}(2L) - 2S_A^{(2)}(L)$ and Fig.~\ref{fig:sm9} for $S_{A}^{(2)}(L) - a L$ using beforehand fitted coefficient $a$. \mhs{For corner cuts, $\scee$ is supposed to cancel out the leading perimeter law automatically, we directly compare the fitting of $\scee$ against $\ln L$ and $1/L$ in Fig.~\ref{fig:sm10}.}

One quantitative way of measuring the quality of the data fitting to a model is through the $\chi^2$ value per degree of freedom, namely $\chi^2/k$. $\chi^2$ is defined as $\chi^2=\sum_{i=1}^M \frac{\left(f(x_{i})-y_{i}\right)^2}{\sigma_i^2 }$, $M$ is the number of data points to be fitted. $k$ denotes the fitting degree of freedom and is obtained by $k=M-r$, where $r$ is the number of fitting parameters and, to be more specific, $r=2$ for the linear fitting functions we used.  $\chi^2/k$ should be typically distributed within the range $[1-\sqrt{2/k},1+\sqrt{2/k}]$. Usually, a large $\chi^2/k$ suggests underfitting, while a small $\chi^2/k$ does not necessarily indicate a satisfactory fitting but can potentially be overfitting or troublesome uncertainties in the data~\cite{Philop,Peter}. For the subtracted EE in Fig.~\ref{fig:sm8}, we have $k=4-2=2$, and a good $\chi^2/k$ lies within $[1-1/\sqrt{1}, 1+1/\sqrt{1}]=[0,2]$. For the scheme in Fig.~\ref{fig:sm9}, $k=9-2=7$ and therefore a reference interval for good $\chi^2/k$ is approximately $[0.46,1.53]$. 

\begin{figure*}[tbp!]
\includegraphics[width=\textwidth]{sm8}
\caption{\textbf{Smooth cuts: subtracted EE versus $\ln L$ and $1/L$, respectively, for different $N$.}
Corresponding $\chi^2/k$ values from linear fits (solid lines) are displayed in each panel. 
%At $N=2,3,5,7$, $\ln L$ fitting has better $\chi^2/k$ values justifying the existence of the subleading log-correction. At $N\gtrsim 8$, $\ln L$ and $1/L$ fitting give comparable $\chi^2/k$ values, manifesting the vanish of the log-correction and permitting the unitary CFT description.
%
For $N \leq 7$, the $\chi^2/k$ values for $\ln L$ fitting are smaller than those for $1/L$ fitting, justifying the existence of a finite subleading log-correction with a finite slope $s$, which is inconsistent with CFTs.
For $N \geq 8$, the $\chi^2/k$ values for $\ln L$ fitting are larger than those for $1/L$ fitting, indicating a vanishing log-correction $s=0$ and a CFT description.
}
\label{fig:sm8}
\end{figure*}

\begin{figure*}[tbp!]
\includegraphics[width=\textwidth]{sm9}
\caption{\textbf{Smooth cuts: direct $S^{(2)}_A-al_A$ versus $\ln L$ and $1/L$, respectively, for different $N$.}
Corresponding $\chi^2/k$ values from linear fits (solid lines) are displayed in each panel. 
The data are consistent with the subtracted EE data shown in Fig.~\ref{fig:sm8}. For $N \leq 7$, the $\chi^2/k$ values for $\ln L$ fitting are smaller than those for $1/L$ fitting, justifying the existence of a finite subleading log-correction. Note that the slope of $S^{(2)}_A-al_A$ versus $\ln L$ corresponds to $-s$, Eq.~\eqref{eq:eqsm2}, in contrast to the slope of the subtracted EE, which corresponds to $+s$, Eq.~\eqref{eq:eqsm3}.
For $N \geq 8$, the $\chi^2/k$ values for $\ln L$ fitting are larger than those for $1/L$ fitting, indicating a vanishing log-correction $s=0$ and a CFT description.
%
%$\chi^2/k$ values are displayed in each case. At $N=2,3,5,7$, $\ln L$ fitting has better $\chi^2/k$ values justifying the existence of the subleading log-correction. At $N\gtrsim 8$, $\ln L$ and $1/L$ fitting give comparable $\chi^2/k$ values, manifesting the vanish of the log-correction and permitting the unitary CFT description.
}
\label{fig:sm9}
\end{figure*}

\begin{figure*}[tb!]
\includegraphics[width=\textwidth]{sm10}
\caption{\mhs{\textbf{Corner cuts: subtracted corner entanglement entropy $\scee$ versus $\ln L$ and $1/L$, respectively, for different $N$.}
Corresponding $\chi^2/k$ values from linear fits (solid lines) are displayed in each panel. 
For $N\leq 7$, $\scee$ exhibits no linear behavior against either $\ln L$ or $1/L$, with large $\chi^2/k$ values. For $N\ge 8$, $\scee$ shows a clear linear behavior against $\ln L$ instead of $1/L$, indicating a finite log-correction $s>0$ from sharp corners, consistent with a CFT description.}}
\label{fig:sm10}
\end{figure*}

\begin{figure*}[tb!]
\includegraphics[width=0.7\textwidth]{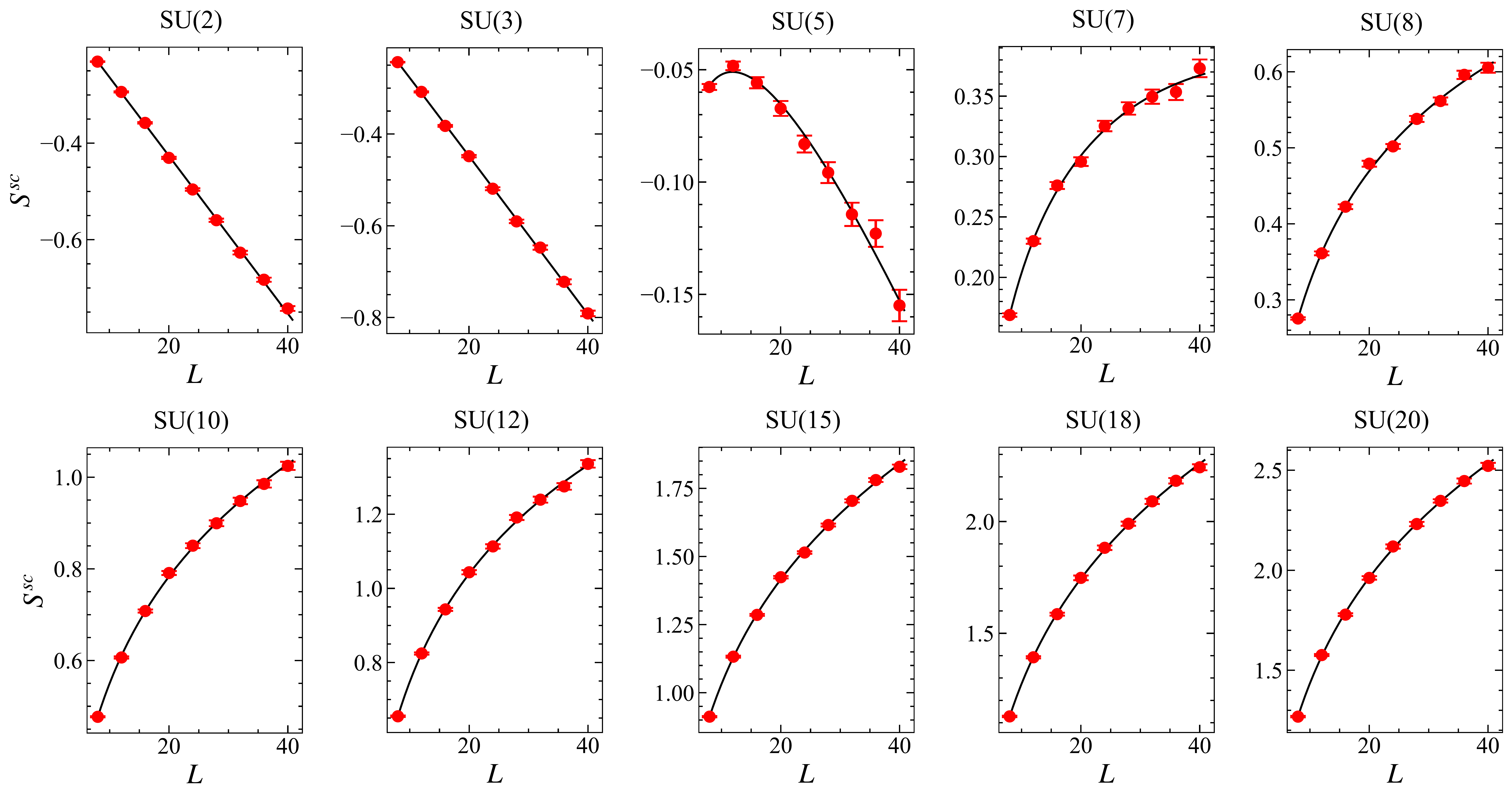}
\caption{\mhs{\textbf{Corner cuts: subtracted corner entanglement entropy $\scee$ versus $L$ for different $N$.}
For $N\leq 7$, the perimeter law coefficients are different in smooth and corner cuts even though both cuts share the same boundary length $l_A=2L$, resulting in the non-linear scaling against $\ln L$ in Fig.~\ref{fig:sm10}. This anomaly is the most obvious in the SU(2) and SU(3) cases where $\scee$ is linear against $L$ reflecting a large remaining perimeter law contribution. The remaining perimeter law contribution decreases when increasing $N$ and vanishes at $N\ge 8$.}}
\label{fig:sm11}
\end{figure*}

%%%%%%%%table begins%%%%%%%%
\begin{table*}[thb]
\caption{Fitted values of log-correction coefficient $s$ in Eq.~\eqref{eq:eq2} for smooth boundaries from direct fitting of $S_A^{(2)} - a l_A$. Values of $s$ that vanish within error bars are colored into grey. Values of $s$ with $L_\text{min}=20$ are omitted for the $N=15$ and $N=20$ cases due to the limited data quality at large $N$.}
\begin{tabular}{c|c|c|c|c|c|c|c|c}
\hline
\hline
\diagbox{$L_\text{min}$}{$N$} & 2         & 3         & 5         & 7 & 8         & 10       & 12       & 15        \\ \hline
8  & -0.272(7) & -0.304(8) & -0.245(9) & -0.20(1) & -0.16(7)                         & -0.11(1)                        & -0.13(2)                        & -0.21(2)                         \\ \hline
12 & -0.28(1)  & -0.34(2)  & -0.24(2)  & -0.19(2) & -0.11(2)                         & -0.04(3)                        & -0.06(4)                        & {\color[HTML]{9B9B9B} -0.05(5)}  \\ \hline
16 & -0.28(3)  & -0.37(4)  & -0.31(5)  & -0.27(5) & {\color[HTML]{9B9B9B} -0.06(6)}  & {\color[HTML]{9B9B9B} -0.05(6)} & {\color[HTML]{9B9B9B} -0.07(8)} & {\color[HTML]{9B9B9B} -0.01(10)} \\ \hline
20 & -0.45(6)  & -0.32(9)  & -0.44(9)  & -0.17(10) & {\color[HTML]{9B9B9B} -0.11(11)} & {\color[HTML]{9B9B9B} 0.01(13)} & -                               & -                                \\ \hline
\hline
\end{tabular}
\label{tab:fitlog_smooth}
\end{table*}

Figures~\ref{fig:sm8} and \ref{fig:sm9} show that for small $N \leq 7$, a linear fit of the subleading correction against $\ln L$ is significantly better than those against $1/L$, as the corresponding $\chi^2/k$ values for the $\ln L$ fitting are consistently smaller than those for the $1/L$ fitting. This shows that these transitions cannot be described by CFTs.
For $N \geq 8$, by contrast, the $\chi^2/k$ values for the $\ln L$ fitting are larger than those for the $1/L$ fitting, indicating a vanishing log-correction $s=0$ for smooth boundaries. Hence, for $N \geq 8$, the DQCPs are consistent with CFT descriptions.
From these results, we conclude that the critical value $\Nc$, above which the log-correction to the EE for smooth boundaries vanishes for all $N$, lies between 7 and 8. The fitted values for $s$ using different $L_\text{min}$ are presented in Table~\ref{tab:fitlog_smooth}, for all cases of $N$ simulated in this work. The cases $N=3, 5, 8, 10, 15$ are plotted in Fig.~\ref{fig:fig2}(c) in the main text.

\mhs{Figure~\ref{fig:sm10} summarizes the fitting analysis for corner cuts with the subtracted corner entanglement entropy, $\scee$. For $N\leq 7$, $\scee$ is not linear against either $\ln L$ or $1/L$ with very large $\chi^2/k$ values. Instead, $\scee$ is linear against $L$ at small $N$s as shown in Fig.~\ref{fig:sm11}, which indicates that the perimeter law coefficient obtained from the smooth cut and corner cut differs despite having the same boundary length $l_A=2L$. This anomaly serves as another evidence for the non-CFT nature of QCPs at $N\leq \Nc$. It may come from the diverse critical fluctuations of the remaining VBS moment in two cuttings.
At $N\ge 8$, the perimeter law coefficient in different cuttings is entirely canceled out which can be seen from the linear behavior of $\scee$ against $\ln L$. This points to a finite log-correction from sharp corners, instead of $1/L$ corrections. The fitted $s$ values at $N\ge \Nc$ are positive and consistent with the prediction of unitary CFTs. }

\end{document}